\documentclass[a4paper,12pt]{article}
\usepackage[utf8]{inputenc}
\usepackage{cancel}
\usepackage{ulem}
\usepackage{amsfonts}
\usepackage{amssymb}
\usepackage{graphicx}
\usepackage{amsmath}
\usepackage{enumerate}
\usepackage{mathtools}
\usepackage{subfloat}
\usepackage{color}
\usepackage{float}
\usepackage{here}
\usepackage{cite}
\usepackage{mathrsfs}
\usepackage{float,epsfig}
\usepackage{dcolumn}

\usepackage{tabularx, ragged2e, booktabs, caption}
\usepackage{blindtext}
\usepackage{caption} 
\usepackage{subcaption}
\usepackage{tablefootnote}
\usepackage{booktabs} 
\usepackage[export]{adjustbox}
\usepackage{authblk}

\usepackage{ulem}

\usepackage{graphicx}
\usepackage{bm}
\usepackage{amsmath,amssymb,amsthm}
\usepackage[colorlinks=true,linkcolor=blue,citecolor=red]{hyperref}
\newcommand{\be}{\begin{equation}}
\newcommand{\ee}{\end{equation}}
\newcommand{\bea}{\setlength\arraycolsep{2pt} \begin{eqnarray}}
\newcommand{\eea}{\end{eqnarray}}

\def\0{{\sst{(0)}}}
\def\1{{\sst{(1)}}}
\def\2{{\sst{(2)}}}
\def\3{{\sst{(3)}}}
\def\4{{\sst{(4)}}}
\def\5{{\sst{(5)}}}
\def\6{{\sst{(6)}}}
\def\7{{\sst{(7)}}}
\def\8{{\sst{(8)}}}
\def\sst#1{{\scriptscriptstyle #1}}

\usepackage{tikz,xcolor,hyperref}

\definecolor{lime}{HTML}{A6CE39}
\DeclareRobustCommand{\orcidicon}{%
	\begin{tikzpicture}
	\draw[lime, fill=lime] (0,0) 
	circle [radius=0.16] 
	node[white] {{\fontfamily{qag}\selectfont \tiny ID}};
	\draw[white, fill=white] (-0.0625,0.095) 
	circle [radius=0.007];
	\end{tikzpicture}
	\hspace{-2mm}
}
\foreach \x in {A, ..., Z}{%
	\expandafter\xdef\csname orcid\x\endcsname{\noexpand\href{https://orcid.org/\csname orcidauthor\x\endcsname}{\noexpand\orcidicon}}
}

\title{\bf Revisiting the Second Law and  Weak Cosmic Censorship Conjecture in High-Dimensional Charged-AdS  Black Hole: an Additional  Assumption}

\author{
 Md Sabir Ali\orcidS{}$^1$\footnote{alimd.sabir3@gmail.com},
 Hasan  El Moumni\orcidH{}$^2$\thanks{h.elmoumni@uiz.ac.ma (Corresponding author)} ,   Jamal  Khalloufi\orcidJ{}$^{2}$\footnote{jamalkhalloufi@gmail.com },
 Karima  Masmar\orcidK{}$^{3}$\footnote{karima.masmar@gmail.com }\footnote{Authors are in alphabetical order.}
\\
{
\small $^{1}$ Department of Physical Sciences, Indian Institute of Science Education and Research Kolkata,
Mohanpur, 741246, India.\\
\small $^{2}$ LPTHE, Physics Department, Faculty of Science,  Ibn Zohr University, Agadir, Morocco. \\
\small $^{3}$ Laboratory of  High Energy Physics and Condensed Matter
HASSAN II University, Faculty of Sciences Ain Chock, Casablanca, Morocco. 
}
}

\vspace{-3.6em}

\date{\today}
\begin{document} 
\maketitle
\begin{abstract}

The verification of the second law of black hole mechanics and the WCCC in the context of enthalpy as mass of the black hole and its related thermodynamic properties has not been tested through a vast number of literature in the recent past. Such studies are of great physical importance as they provide us with a large number of information regarding the thermodynamics and the dynamics of AdS black hole systems. We invest the prior limited surveys of such analysis to investigate the WCCC for the $D$- dimensional asymptotically AdS-charged black holes characterized by its mass ($M$), electric charge ($Q$), and AdS radius ($l$) under the absorption of  scalar particles of charge $q$. We examine the WCCC by analyzing the energy-momentum condition of the electrically charged particles as absorbed by the black holes. We prove that the conjecture is well verified irrespective of whether the initial black hole configurations are extremal or non-extremal by changing its charge, the AdS radius, and their variations. We show that the first law and the WCCC are valid for all spacetime dimensions ($D$) independent of the choice of the parameters characterizing the black holes. But to verify the second law in the extremal and non-extremal configurations one has to be very cautious as it gets strongly affected by the choices of the values of the black hole parameters and their variations. 
  In other words, we use charged particle dynamics as described by the Hamilton-Jacobi equation to obtain the energy-momentum relation as the charged particle dropped into the higher dimensional charged AdS black hole and verify the thermodynamic laws when the scalar charged particle gets absorbed by the black holes 
  and correspondingly the black hole neutralization in different manners. Additionally, we further probe the validity of 
  WCCC in such a black hole background. 
   In the context of the extended phase space, taking the grand canonical potential into account allow us to obtain the missing information about the variation of the cosmological constant necessary to construct the extended phase space, namely the notion of the black hole pressure, and which is absent in the previous literature so far.

{\noindent}

\end{abstract}
\tableofcontents

\section{Introduction}

Because of the seminal work of Hawking \cite{Hawking:1976de}, we realize that a black hole isn't just a celestial object, but yet,  a thermodynamic system. Just like the ordinary thermodynamic systems, there are additionally, four laws for the black hole thermodynamic framework \cite{Henneaux:1985tv}, which are valid near the horizon. The significance of black hole thermodynamics comes from the discovery of a  fascinating connection between theories of gravity, and quantum ﬁeld theory. This bridge allows us to construct the first brick in our understanding of a possible quantum gravity theory.

Furthermore, the event horizon where the laws of the black hole thermodynamics are defined, disguises the curvature singularity from the outside observers located at infinity \cite{Penrose:1964wq}. When the horizon is destroyed, the singularity turns into what is called a naked singularity. At such a location, any physical law is no longer valid and it's broken \cite{Penrose:1969pc}. Consequently, the horizon is assumed to be stable according to the weak cosmic censorship conjecture  (WCCC). Despite the fact that it is generally accepted without demonstration procedure that this conjecture applies to black holes, its validity should be tested.  The validity test can be realized by adding an electrically charged particle having its own energy-momentum and is getting absorbed by the black hole \cite{WALD1974548}. As a result, a catastrophic countdown may happen to the event horizon as the infalling charged particles are absorbed, the singularity may disappear and it can become the naked one indicating the violation of the WCCC. If the horizon survives to exist, it should hide the singularity behind and the WCCC is respected.

A lot of investigations regarding the enthalpy formalism
have been proposed in the recent past in order to sufficiently probe the quantum gravitational nature of the AdS black holes\cite{Dolan:2011xt,Kubiznak:2012wp,Gunasekaran:2012dq,Hennigar:2015esa,Belhaj:2015hha,Xu:2014kwa,Chabab:2017knz,Wang:2018xdz,Luo:2022gss}. Such studies have been emphasized and limited to the first law and dimensional verification of the Smarr-Gibbs-Duhem relation. However, the propositions regarding the thermodynamics laws at the black hole horizon and the WCCC are still lacking concrete analysis for the AdS spacetime. The first two laws of the black hole mechanics and also the WCCC were studied for the charged AdS black holes\cite{Gwak:2017kkt} and with an additional scalar hair \cite{Chen:2019nsr}. The first law and the WCCC are not violated for such cases, whereas the second law is found to be violated in the near extremal case although it is held for the extremal cases for particular choices of the charge parameter of the electrically charged particles that are absorbed by the black holes. Such studies have also been extended to the Born-Infeld-AdS\cite{art7}, low-dimensions black hole models\cite{Zeng:2019huf}, and the Kerr-Newman-AdS black holes. In the Kerr-Newman-AdS black holes too\cite{Gwak:2018akg} we have the WCCC to be valid and the second law might be valid or violated if the charge parameter or the spin parameter are varied. As a limiting case when the charge parameter $Q=0$, we can refer our reader to \cite{Zeng:2019aao}, for a similar analysis of the Kerr-AdS black holes. More interestingly, the WCCC is found not to be held in the dimensionally extended near-extremal AdS black holes in $f(R)$ gravity \cite{He:2019kws}.

Concretely speaking, numerous research activities have been performed to show that the WCCC might still be plausible for a variety of black hole configurations. As a consequence of such analysis, people also examined whether the first two laws of black hole thermodynamics are valid when the particle absorption process takes place. Although a lot of work has been done, no consistent conclusion has been reached on a concrete basis, see for example, refs.~\cite{Chen:2018yah,He:2019kws,Zeng:2019aao,Hu:2019zxr,Chen:2019nhv,Yang:2020czk,Feng:2020tyc,Gwak:2019asi,Liang:2020hjz,Li:2020nnz,Nie:2021rhz,bulk,Wang:2019jzz,Ahmed:2022dpu,Yang:2022yvq}. In particular, while studying the viability of the laws of thermodynamics and/or to ensure existence of the stable horizon via particle absorption, two assumptions associated with the energy of the particle have been found in the literature. The first one is due to Gwak's work \cite{art9},  where it is assumed that the changes in the black hole's internal energy happened due to the change in the infalling particles, i.e., $E = dU$. Under this assumption, even though the first law is satisﬁed, the second law is found to be violated. Moreover, the extremality and the near-extremality conditions of the black holes remain the same. The second assumption\cite{Hu:2019lcy} stipulates that the energy of the absorbed particle is accompanied by the change in black hole enthalpy, $E = dM$. Under such assumptions, the first, as well as the second laws of the black hole mechanics, are identically satisfied. Besides, black hole's event horizon is stable for the extremal as well as the near-extremal configurations. But at the core of this proposal lies  a deficiency of the argument that one can think off:  the assumption that in the conventional holographic dictionary one deals with the large $N$ regime of the boundary CFT, but the value of $N$, however large it is, taken as a constant entity which is not in favor of the formalism in which the cosmological constant is considered no more as a constant quantity but as a variable thereby contributing to the first law \cite{Henneaux:1984ji,Kubiznak:2012wp,art20,Wei:2015iwa}. Furthermore, associating the connection of the holographic description to the extended phase space has led us to find a variety of rich phase structure of the AdS black holes\cite{Zhang:2014uoa,Belhaj:2015uwa,Chabab:2015ytz}, which has further revealed a new notion to study the heat engines from holographic perspectives \cite{Johnson:2014yja}. Such findings in the phase structure has unveiled a new door, that is known in the scientific community as the AdS/CFT duality. This way, we had been able to incorporate the grand canonical ensemble, where we restored the missing piece of information by varying the cosmological constant and its conjugate fluid dynamical pressure. Furthermore, no evidence of second law violation is no longer valid in the extended black hole thermodynamics\cite{nous}.\\

The outline of the paper is as follows. In Section 2, we introduce the motion of a test electrically charged particle in a charged AdS black hole background in the $D$-dimensional and some of the thermodynamic quantities are briefly presented along with the fluid dynamical cosmological constant. In section 3, we derive the first law of thermodynamics when an electrically charged particle gets absorbed and put the second law under examination for different energy assumptions, namely: internal energy, enthalpy, and the grand canonical potential. We prove by incorporating the last quantity to show the second law to be held, although the cosmological constant varies. Furthermore, within section 4 and by recalling the minimum point of the blackening factor $f(r)$ and probing the variation around it we assert that it should respect the WCCC even though we varied the cosmological constant for such a black hole background. Consequently, one discusses the second law and the neutralization process at constant black hole’s event horizon entropy, by considering the effective entropy formalism and then an isobaric process. To complete our study, in section 5, we turn our attention to a grand canonical description within a fixed grand canonical potential absorption, and thereby permitting an elegant check to the third law even though we include the $(P,V)$ pairs in first law. The last section is devoted to the conclusion.



%
%
%
%

\section{On a charged particle dynamics  in the  Charged Anti-de-Sitter black hole background}
In this section, we would investigate the motion of a charged particle to test the laws of thermodynamics, especially the second law, and the WCCC in the high dimensional charged AdS background.
\subsection{The charged Anti-de-Sitter Black Hole thermodynamics: A succinct review of the extended phase space}
To start with we write the Einstein-Maxwell action in the $D$-dimensional AdS spacetime 
\begin{equation}\label{1}
\mathcal{S} = -\dfrac{1}{16 \pi} \int d^Dx\sqrt{-g} (R - \mathcal{F}_{\mu \nu}\mathcal{F}^{\mu \nu} - 2 \Lambda).
\end{equation}
In which $R$ is the Ricci  scalar, the Maxwell field strength $\mathcal{F}_{\mu \nu}$  and electric potential $A_\mu$ are defined to be
\begin{equation}\label{2}
\mathcal{F}_{\mu \nu} = \partial_\mu A_\nu - \partial_\nu A_\mu, \quad \quad A_\mu\;dx^{\mu}= - \dfrac{Q_D}{r^{D-3}} dt.
\end{equation}
The metric pertaining to the above action, for a static spherically symmetric AdS spacetime in higher dimensions is written as
\begin{equation}\label{3}
ds^2 = - f(r) dt^2 + \dfrac{dr^2}{f(r)}+ r^2 d\Omega_{D-2}^2,
\end{equation}
where the blackening factor is 
\begin{equation}\label{4}
f(r) = 1 - \dfrac{2 M_D}{r^{D-3}} + \dfrac{Q_D^2}{r^{2D-6}} + \dfrac{r^2}{l^2},
\end{equation}
and $d\Omega_{D-2}^2$ represents line element of the $(D-2)$-dimensional unit sphere and is denoted as
\begin{equation}\label{5}
 d\Omega_{D-2} = \sum_{i=1}^{D-2}\left( \prod_{j=1}^{i} \sin^2\theta_{j-1}\right) d \theta_i, \quad \theta_0 = \dfrac{\pi}{2}, \quad \theta_{D-2} = \phi .
\end{equation}
The metric components are functions of mass and charge parameters $M_D$ and $Q_D$ along with an AdS radius $l$. 
These parameters are proportional to the black hole mass $M$, electric charge $Q$, and cosmological constant $\Lambda$ \cite{art1}
\begin{equation}\label{6}
M = \dfrac{\left( D-2 \right)\Omega_{D-2} }{8 \pi} M_D, \quad Q = \dfrac{\left( D-2 \right)\Omega_{D-2} }{8 \pi} Q_D, \quad \Lambda = - \dfrac{\left( D-1\right) \left( D-2\right) }{2 l^2}.
\end{equation}
The RN-AdS black hole has an event horizon at $r=r_h$ where the constraint  $f(r_h)=0$ is satisfied. In terms of $r_h$, the Hawking temperature $T$, Bekenstein-Hawking entropy $S$, and the electric potential are expressed, respectively, as
\begin{equation}\label{7}
T =\frac{f'(r_h)}{4\pi}= \dfrac{1}{2 \pi l^2} \left( r_h - \dfrac{\left( D-3\right)  Q_D^2 l^2}{r_h^{2 D-5 }} + \dfrac{\left( D-3\right)  M_D l^2}{r_h^{ D-2 }} \right) , \quad S = \dfrac{\Omega_{D-2}r_h^{D-2}}{4}, \quad \Phi = \dfrac{Q_D}{r_h^{D-3}}.
\end{equation}

In the black hole's enthalpy formalism, the cosmological constant, $\Lambda$, plays the role of pressure $P$, and its conjugate quantity is viewed as a thermodynamic volume, $V$, of the black hole. The connection of the thermodynamic pressure and volume in $D$-dimensional AdS spacetime, respectively, read \cite{art2}
\begin{equation}\label{8}
P = - \dfrac{\Lambda}{8\pi} = \dfrac{\left( D-1\right) \left( D-2\right) }{16 \pi l^2}, \quad V = \dfrac{\Omega_{D-2}}{D-1}r_h^{D-1}.
\end{equation}
In thermodynamics of black holes, the first law states that the infinitesimal change of the masses between two nearby solutions equals the differences of entropy times the Hawking temperature plus some additional work terms \cite{art2, art3}. When the pressure term is included we obtain a generalization of the ﬁrst law of black hole
mechanics. Thus the first law for a $U(1)$ charged AdS black holes with the inclusion of $(P,V)$ pair yields 
\begin{equation}\label{9}
dM = T dS + V dP + \Phi dQ.
\end{equation}
Remember that when the pressure term is included in the laws of thermodynamics, the black hole mass $M$ should play the role of enthalpy \cite{ar4,art5}. All the essential thermodynamics quantities in hand, we shall intend, in the next section to obtain the energy-momentum relation of a charged particle near the event horizon. It is noticeable that the energy-momentum of the charged particle is absorbed by the high dimensional charge AdS black hole. We are interested in a scalar particle only, so we employ the following Hamilton-Jacobi equation to probe the dynamical nature of the absorbed charged particle.
\subsection{An infalling particle in a black hole: Energy and momentum}
In this section, we consider the dynamics of an electrically charged particle whose energy-momentum is controlled by the charge black hole spacetime. To deal with the dynamics of an electrically charged particle we find out the connections among the conserved quantities, such as the conserved energy and the angular momentum.

We construct the equation of motion of the electrically charged particle with charge $q$, using the Hamilton-Jacobi formalism in the presence of the gauge potential $A_\mu$
\begin{equation}\label{10}
g^{\mu \nu} \left( p_\mu - q A_\mu \right) \left( p_\nu - q A_\nu \right) = - m^2,
\end{equation}
where $m$ is the rest mass of the particle, and $p_\mu$ is the four momentum of the charged scalar particle with 
\begin{equation}\label{11}
p_\mu = \partial_{\mu}I,
\end{equation}
an where the quantity $I$ stands for the Hamilton-Jacobi action. As the spacetime admits two Killing vectors corresponding to $t$ and $\phi$, one can write the Hamilton-Jacobi action using the variable separations
\begin{equation}\label{12}
I = -E t - I_r(r)  + L \phi + \sum_{i=1}^{D-3} I_{\theta_i}(\theta_i).
\end{equation}
Herein, the conserved quantities energy ($E$) and angular momentum ($L$), respectively, arise from the time translation symmetry along $t$-axis and the rotation symmetry about $\theta_{D-2}=\phi$-axis.   For velocities of the particle  Eq.\eqref{3} we can express the Hamilton-Jacobi equation using Eq.\eqref{12} as
 \begin{equation}\label{13}
 m^2 - \dfrac{1}{f(r)}\left( E + q A\right)^2 + f(r) \left( \partial_r I_r(r) \right)^2 + \dfrac{1}{r^2} \sum_{i=1}^{D-3}\left( \prod_{j=1}^{i} \sin^{-2}\theta_{j-1}\right) \left( \partial_{\theta_i} I_{\theta_i}({\theta_i}) \right)^2 + \dfrac{1}{r^2} \left( \prod_{j=1}^{D-2} \sin^{-2}\theta_{j-1}\right) L^2 = 0 .
 \end{equation}
 Separating the angular part from the radial one gives
 \begin{equation}\label{14}
\sum_{i=1}^{D-3}\left( \prod_{j=1}^{i} \sin^{-2}\theta_{j-1}\right) \left( \partial_{\theta_i} I_{\theta_i}({\theta_i}) \right)^2 + \left( \prod_{j=1}^{D-2} \sin^{-2}\theta_{j-1}\right) L^2 = C_1 ,
 \end{equation} 
 where $C_1$ is a constant such that
 \begin{equation}\label{15}
 C_1 =  \dfrac{r^2}{f(r)}\left( E + q A\right)^2 - m^2 r^2  + r^2 f(r) \left( \partial_r I_r(r) \right)^2  .
 \end{equation}
wherefore,  Eq.\eqref{12} can be written as
 \begin{equation}\label{16}
 I = - E t + \int dr \sqrt{R}  + L \phi +  \sum_{i=1}^{D-3} \int d\theta_i \sqrt{\Theta_i },
 \end{equation}
 in which, we have set 
 \begin{equation}\label{17}
 \begin{split}
 R &= \dfrac{\left( E + q A\right)^2}{f(r)^2}\left[ 1 - \dfrac{\left( C_1 + m^2 r^2\right) f(r)}{\left( E + q A\right)^2 r^2}\right], \\
 \Theta_i &= C_i - \dfrac{C_{i+1}}{\sin^2 \theta_i}, \quad C_{D-2} = L^2 .
 \end{split}
 \end{equation}
 Hence, the radial momentum $p^r$ can be written as
 \begin{equation}\label{18}
 p^r = \pm \left( E + q A\right) \sqrt{1 - \dfrac{\left( C + m^2 r^2\right) f(r)}{\left( E + q A\right)^2 r^2}}.
 \end{equation}
We shall investigate how a charged scalar particle is engulfed by a black hole and how it affects the black hole's thermodynamics. We focus on how the particle behaves when it is close to the event horizon, where $f(r)\to 0$. Therefore, the energy of the particle reads
 \begin{equation}\label{19}
 E = \left| p^r_h\right| + \dfrac{q Q_D}{r_h^{D-3}}.
 \end{equation}
 It is noteworthy to mention that we have chosen the positive sign in front of $\left| p^r_h\right|$ to keep $E$ and $\left| p^r_h\right|$ are of the same signs and also to ensure the flow of time in the positive direction \cite{art6}.
After obtaining the energy-momentum relation, we shall employ it to check the thermodynamic laws and the WCCC in the black hole enthalpy formalism.
\section{The second law of thermodynamics under different considerations}
The energy and charge should be conserved when the charged particle falls into the black hole, and it is given by the sum of the electric potential of the charged particle and its radial momentum near the event horizon. We use this relation in order to incorporate the thermodynamics of the charged AdS black holes. Such a phenomenon occurs when the ingoing charged scalar particles with energy $E$ and electric charge $q$ are absorbed by the charged AdS black holes. Consequently, both the black hole's  thermodynamic potential and the charge are changed with the horizon radius and the pressure. In such a situation, the mass, charge, horizon radius, and pressure of the black holes should also change. This leads us to connect the changes in the energy and the electric charge of the particle to the first law.

Rigorously speaking,  we will look for which thermodynamic potential is varied by the same quantity as that of the particle, and this potential's variation is supported by the change of the black hole configurations following in view of the first law.
\subsection{The first assumption: internal energy }
\begin{figure}[H]
	\centering 
	\begin{subfigure}[h]{0.45\textwidth}
		\centering \includegraphics[scale=0.5]{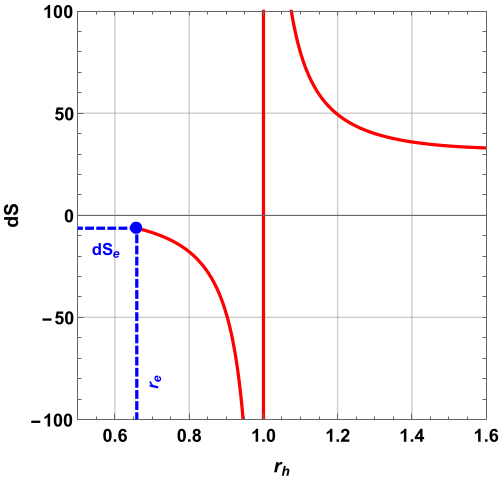}
		\caption{$D=4$}
		\label{f1_1}
	\end{subfigure}
	\hspace{1pt}	
	\begin{subfigure}[h]{0.45\textwidth}
		\centering \includegraphics[scale=0.5]{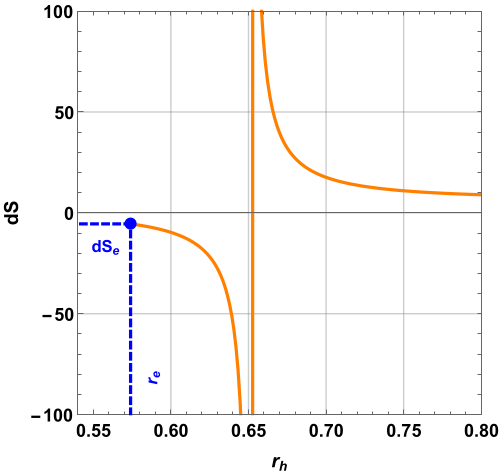}
		\caption{$D=5$}
		\label{f1_2}	
	\end{subfigure}
	\hspace{1pt}	
\begin{subfigure}[h]{0.45\textwidth}
	\centering \includegraphics[scale=0.5]{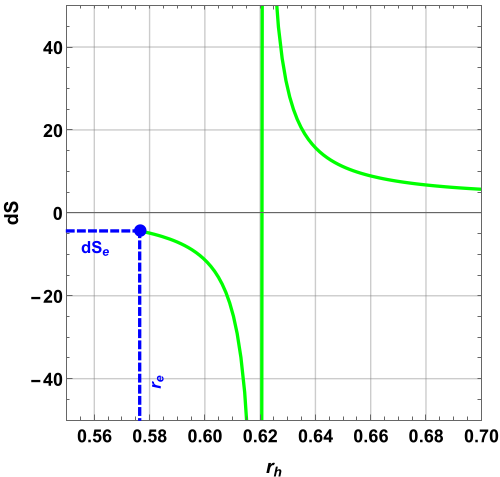}
	\caption{$D=6$}
	\label{f1_3}	
\end{subfigure}
	\hspace{1pt}	
\begin{subfigure}[h]{0.45\textwidth}
	\centering \includegraphics[scale=0.5]{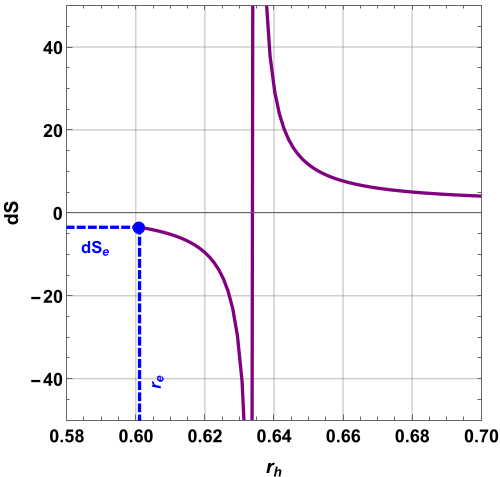}
	\caption{$D=7$}
	\label{f1_4}	
\end{subfigure}
	
	\caption{\footnotesize\it Entropy variation $dS$ as a function of event horizon radius $r_h$ for different dimensions  with $\left| p^r_h\right| = 1$, $Q=1$, $l=1$ and $dQ = 0.1$. }
	\label{f1}
\end{figure}

We presume that the energy of the particle alters the black hole's internal energy \cite{art7, art8, art9}.
Given as $U\left( Q, S, V \right)$, the internal energy depends on the charge, entropy, and volume of the black hole, 
and their variations as the charged particle is getting absorbed as it falls into the black hole. It should also be  $dQ$, $dS$, and $dV$ dependent, as can be shown from the following equation. 
\begin{equation}\label{20}
E = dU = d\left( M-PV\right) , \quad \quad dQ = q.
\end{equation}
From the first law of thermodynamics given in \eqref{13}, we have
\begin{equation}\label{21}
dU = T dS -  P dV + \Phi dQ. 
\end{equation}
Thus the variation of the internal energy in the limit of $f(r)\to 0$, is obtained from Eq.\eqref{9} as
\begin{equation}\label{22}
dU = \left| p^r_h\right| + \dfrac{q Q_D}{r_h^{D-3}}.
\end{equation}
The entropy changes accordingly as 
\begin{equation}\label{23}
dS = \left( D-2\right) S \dfrac{d r_h}{r_h},
\end{equation}
where the change of the black hole event horizon, $dr_h$, is rewritten as independent variables just like $(dQ, \left| p^r_h\right|)$ pairs of the particle. However, near the horizon, $df(r_h) = df_h$ does not change since $f(r_h + dr_h) = 0$, that is to say
\begin{equation}\label{24}
df_h = \dfrac{\partial f_h}{\partial r_h} dr_h + \dfrac{\partial f_h}{\partial M} dM + \dfrac{\partial f_h}{\partial Q} dQ + \dfrac{\partial f_h}{\partial l} dl  = 0.
\end{equation}
We note from \eqref{7} and \eqref{8} that the entropy and volume depend only on the event horizon radius and that their changes are proportional to $dr_h$, however, the variation of internal energy depends on the variation of the volume and entropy differently, as we 
Furthermore, the event horizon radius fluctuation $d r_h$ is independent of $dl$, and the first principle of thermodynamics does not account for pressure change.
Thus increasing the internal dose does not imply the increase of the event horizon and hence the entropy!
We calculate the variation of the event horizon while simultaneously resolving \eqref{21}, \eqref{22}, and eqref{24} in order to confirm this prediction.
Consequently, the event horizon's fluctuation is obtained to be
\begin{equation}\label{25}
d r_h = \dfrac{16 \pi \left| p^r_h\right| r_h^3}{\left( D-2\right) \left( D-3\right) \left( 1-Q_D^2 r^{6-2 D}\right) \Omega_{D-2} r^{D-1} }.
\end{equation}
Using \eqref{7} and \eqref{8}, we can write \eqref{25} as 
\begin{equation}\label{26}
d r_h = \dfrac{4 \pi \left| p^r_h\right| r_h^2}{\left( D-2\right) \left( D-3\right) S \left( 1-\Phi^2\right) }.
\end{equation}
Therefore, the corresponding variation in the the black hole entropy is expressed as
\begin{equation}\label{27}
d S = \dfrac{4 \pi \left| p^r_h\right| r_h}{\left( D-3\right) \left( 1-\Phi^2\right) }.
\end{equation}
When $\Phi^2 > 1 $, the fluctuation of entropy could be negative. Therefore, when $r_h< r_0$, we have a violation of the second rule of thermodynamics, whit
\begin{equation}\label{28}
r_0 = Q_D ^{\dfrac{1}{D-3}}.
\end{equation}
We plot in Fig.\ref{f1} the variation of the black hole entropy as a function of event horizon radius with different dimensions.
 We observe that the second law is violated particularly for the extremal condition  where $dS_e< 0$ and for all black hole configurations when $r_h<r_0$. Whereas, for $r_h>r_0$ the entropy varies positively and the second law is respected. Moreover, the variation of the entropy diverges at $r_h = r_0$.

\begin{figure}[H]
	\centering \includegraphics[scale=0.5]{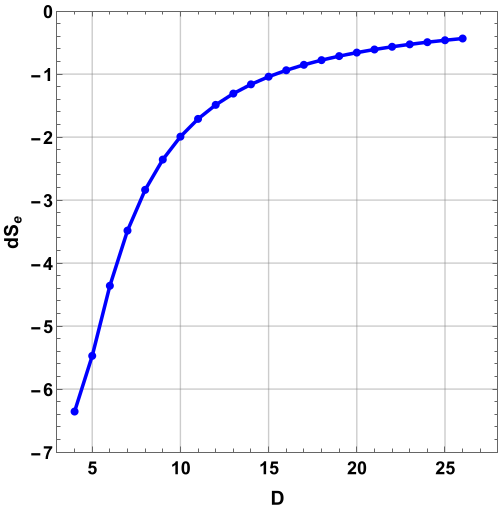}
	\caption{\footnotesize\it Variation of entropy $dS_e$ for the extrema black hole as a function of spacetime dimension $D$, with $\left| p^r_h\right| = 1$, $Q=1$, $l=1$ and $dQ = 0.1$. }
	\label{f2}
\end{figure}

The variation of the extremal black hole's entropy is obtained from \eqref{26} using the expression of the extremal value of the electric charge :
\begin{equation}\label{29}
dS_e = - \dfrac{4 \pi \left| p^r_h\right| l^2}{\left( D-1\right) r_h  },
\end{equation}
which is always negative for all dimensions as we can see in Fig.\ref{f2} 
where we have plotted the entropy variations in the extremal case as a function of the spacetime dimension.

 This behavior is unphysical because the black hole is regular where $r_h = r_0$ and hence the divergence of the entropy does not make sense. Therefore, we conclude that the energy of the particle does not have absorbed via the internal energy of the black hole.

\subsection{The second assumption: enthalpy }

In this subsection, we suppose that the energy of the particle changes the enthalpy $H$ of the black hole as it was assumed in \cite{art10}.  The enthalpy is a Legendre transformation of the internal energy such as
\begin{equation}\label{30}
H = U+ PV.
\end{equation}
In the present assumption, the enthalpy $H$ relates to the black hole's mass $M$.
Given as $H\left( Q, S, P \right)) $, the enthalpy is a function of the black hole's charge, entropy, and pressure. Thus, according to the following equation, the variation in enthalpy when the charged scalar particle falls into the black hole should be written in terms of $dQ$, $dS$, and $dP$
\begin{equation}\label{31}
E = dH = dM , \quad \quad dQ = q.
\end{equation}
Thus the variation of the enthalpy is given from Eq.\eqref{19}
\begin{equation}\label{32}
dM = \left| p^r_h\right| + \dfrac{q Q_D}{r_h^{D-3}}.
\end{equation}
According to \eqref{9}, the variation of the black hole mass reads 
\begin{equation}\label{33}
dM = \dfrac{\partial M}{\partial r_h} dr_h + \dfrac{\partial M}{\partial Q} dQ + \dfrac{\partial M}{\partial l} dl.
\end{equation}
Hence, the change in the event horizon is given by
\begin{equation}\label{34}
dr_h = \dfrac{2 r_h^2 \left[ 2 \pi \left| p^r_h\right| l^3 + \left( D-2\right) r_h S dl\right]  }{\left( D-2\right) S l \left[ \left( D-1\right) r_h^2 + \left( D-3\right) l^2 \left( 1-\Phi^2\right) \right] },
\end{equation}
which verifies the expression given in \cite{art10} for four dimensions. Therefore, the differential change in the black hole's entropy is rewritten as
\begin{equation}\label{35}
d S = \dfrac{2 r_h \left[ 2 \pi \left| p^r_h\right| l^3 +  \left( D-2\right) r S dl\right]  }{  l \left[ \left( D-1\right) r_h^2 + \left( D-3\right) l^2 \left( 1-\Phi^2\right) \right] }
\end{equation}
Our problem is that we have any information about how the pressure varies. In other words, we have two unknown quantities, $dr_h$ and $dl$, and we have just one equation which is the first law of thermodynamics. In fact, the equation \eqref{24} is identical to the equation for the first law and provides no further details.
Therefore, in order to close the system, we require an additional dynamical equation. 

\subsection{The third assumption: grand potential}

The partition function is crucial whenever we need statistical and thermodynamical descriptions of a system. 
In the semiclassical limit, we have 
\begin{equation}\label{36}
\mathcal{Z} = e^{- \mathcal{S}_E};
\end{equation}
where $\mathcal{S}_E$ is the euclidean action which is related to the free energy $F$ in the grand canonical ensemble by \footnote{In usual thermodynamical systems we have $F - N \mu = T \mathcal{S}_E$  where $F$ is Helmholtz energy, $N$ is the number of particles and $\mu$ is the chemical potential. In black hole thermodynamics, we replace Helmholtz energy, $U-TS$, with Gibbs energy, $M -T S$, because the black hole mass plays the role of enthalpy.}
\begin{equation}\label{37}
F - Q \Phi = - T \log \left( \mathcal{Z}\right)  = T \mathcal{S}_E,
\end{equation}
such that $F - Q \Phi $ is the grand potential \cite{art11, art12}. The least action principle in black hole thermodynamics is read \cite{Wang:2018xdz,art14,art15}, 
\begin{equation}\label{38}
d\left( F - Q \Phi\right)  = 0,
\end{equation}
traducing the black hole's dynamical stability and it's ensured by minimizing potential. 

Now that we are in possession of an additional equation in addition to the first law and we can solve our problem correctly.
When \eqref{32} and \eqref{38} are resolved concurrently, the event horizon and AdS radius vary as follows
\begin{equation}\label{39}
\begin{split}
dr_h &= \dfrac{2 \pi E r_h^2}{\left( D-2\right) \left( D-3\right) S}, \\
dl &= -\dfrac{l}{2 P} \left[ \dfrac{2 \left| p^r_h\right| + E \left( \Phi^2 -1\right)  }{2 V} - \dfrac{8 \pi P E r_h^2}{\left( D-2\right) \left( D-3\right) V}\right].
\end{split}
\end{equation}   
Therefore, the modification to the black hole's entropic change is re-expressed as
\begin{equation}\label{40}
d S = \dfrac{2 \pi E r_h}{ \left( D-3\right) }, 
\end{equation}
thanks to \eqref{8}, one can write down the variation of the pressure as  
\begin{equation}\label{41}
dP = - 2 P \dfrac{dl}{l},
\end{equation}
thus
\begin{equation}\label{42}
dP = \dfrac{2 \left| p^r_h\right| + E \left( \Phi^2 -1\right)  }{2 V} - \dfrac{8 \pi P E r_h^2}{\left( D-2\right) \left( D-3\right) V}.
\end{equation}

\begin{figure}[H]
	\centering 
	\begin{subfigure}[h]{0.45\textwidth}
		\centering \includegraphics[scale=0.4]{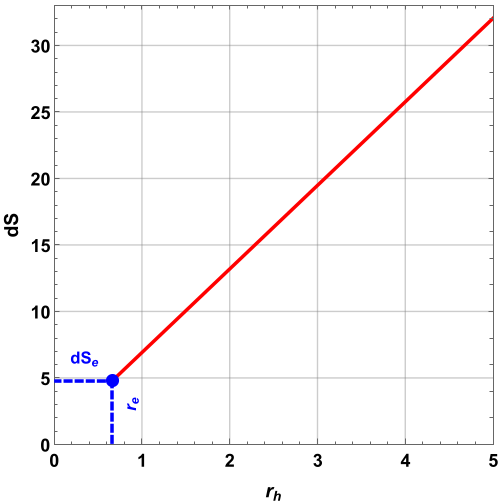}
		\caption{$D=4$}
		\label{f3_1}
	\end{subfigure}
	\hspace{1pt}	
	\begin{subfigure}[h]{0.45\textwidth}
		\centering \includegraphics[scale=0.4]{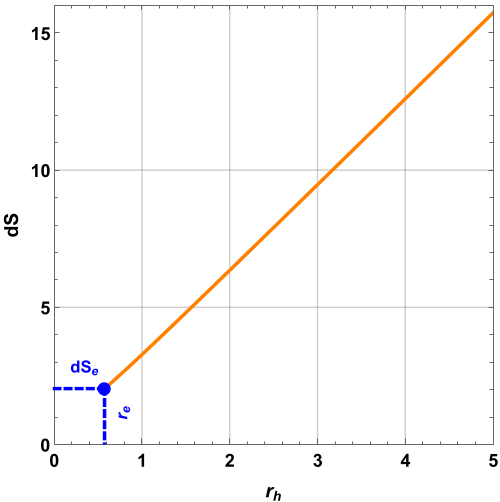}
		\caption{$D=5$}
		\label{f3_2}	
	\end{subfigure}
	\hspace{1pt}	
	\begin{subfigure}[h]{0.45\textwidth}
		\centering \includegraphics[scale=0.4]{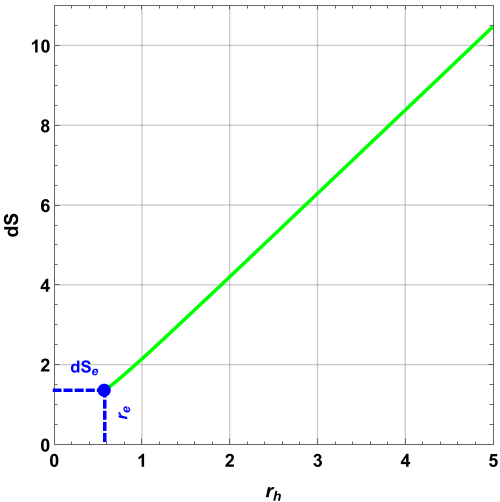}
		\caption{$D=6$}
		\label{f3_3}	
	\end{subfigure}
	\hspace{1pt}	
	\begin{subfigure}[h]{0.45\textwidth}
		\centering \includegraphics[scale=0.4]{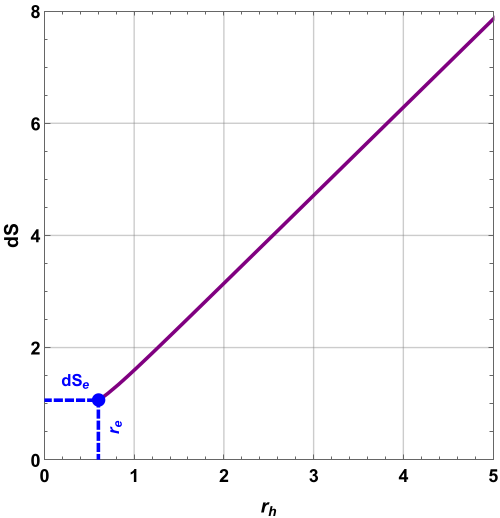}
		\caption{$D=7$}
		\label{f3_4}	
	\end{subfigure}
	
	\caption{\footnotesize\it Entropy variation $dS$ as a function of event horizon radius $r_h$ for different dimensions  with $\left| p^r_h\right| = 1$, $Q=1$, $l=1$ and $dQ = 0.1$. }
	\label{f3}
\end{figure}

We have, from \eqref{40}, the variation of the entropy is always positive and proportional to event horizon radius at a fixed absorbed energy for all dimensions \footnote{The energy of the absorbed particle depends on $r_h$ as we can see in \eqref{19}, but it tends to $\left| p^r_h\right|$ when $r_h \to \infty$ or $q\to 0.$}. We show the variation of black hole entropy in Fig. \ref{f3} as a function of the event horizon radius for various AdS spacetime dimensions, and we note that the variation of entropy is positive and increases linearly as the black hole gets bigger.

\begin{figure}[H]
	\centering 
	\begin{subfigure}[h]{0.45\textwidth}
		\centering \includegraphics[scale=0.4]{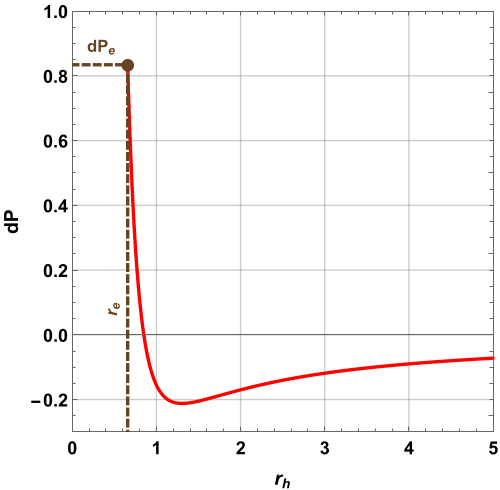}
		\caption{$D=4$}
		\label{f4_1}
	\end{subfigure}
	\hspace{1pt}	
	\begin{subfigure}[h]{0.45\textwidth}
		\centering \includegraphics[scale=0.4]{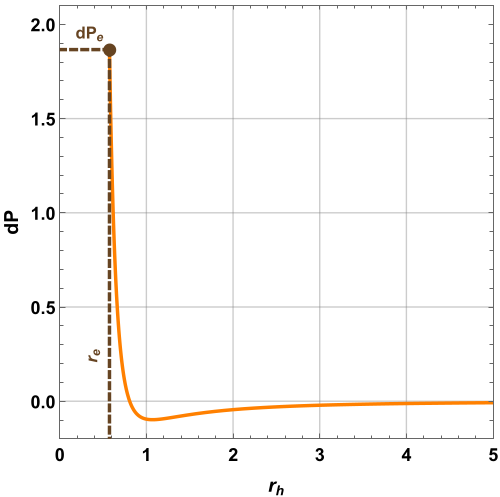}
		\caption{$D=5$}
		\label{f4_2}	
	\end{subfigure}
	\hspace{1pt}	
	\begin{subfigure}[h]{0.45\textwidth}
		\centering \includegraphics[scale=0.4]{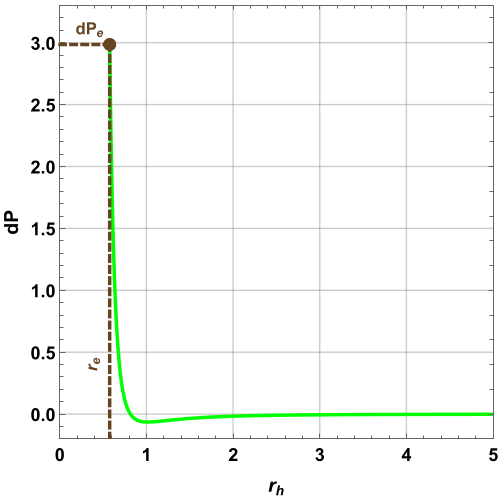}
		\caption{$D=6$}
		\label{f4_3}	
	\end{subfigure}
	\hspace{1pt}	
	\begin{subfigure}[h]{0.45\textwidth}
		\centering \includegraphics[scale=0.4]{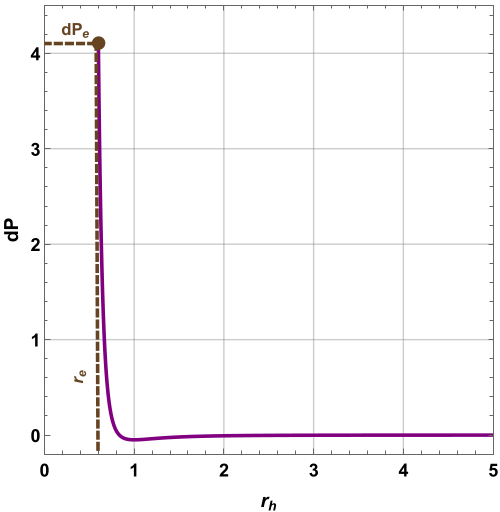}
		\caption{$D=7$}
		\label{f4_4}	
	\end{subfigure}
	
	\caption{\footnotesize\it Pressure variation $dP$ as a function of event horizon radius $r_h$ for different dimensions  with $\left| p^r_h\right| = 1$, $Q=1$, $l=1$ and $dQ = 0.1$. }
	\label{f4}
\end{figure}

We plot in Fig.~\ref{f4} the variation of the black hole pressure as a function of the event horizon radius for different AdS spacetime dimensions. We observe that the differential change in the pressure is always positive for the extremal black hole and it becomes quickly negative for the non-extremal case and vanishes when the black hole gets larger. Hence, for the black hole with large horizon radius, we can consider the absorption of a particle as an isobar thermodynamic process, particularly for higher dimensional black holes. 

\begin{figure}[!t]
	\centering 
	\begin{subfigure}[h]{0.45\textwidth}
		\centering \includegraphics[scale=0.4]{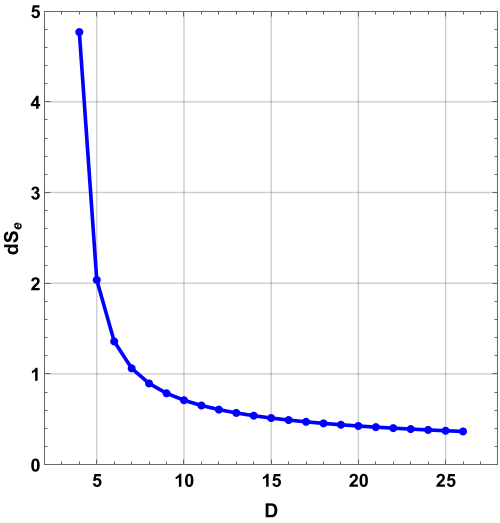}
		\caption{}
		\label{f5_1}
	\end{subfigure}
	\hspace{1pt}	
	\begin{subfigure}[h]{0.45\textwidth}
		\centering \includegraphics[scale=0.4]{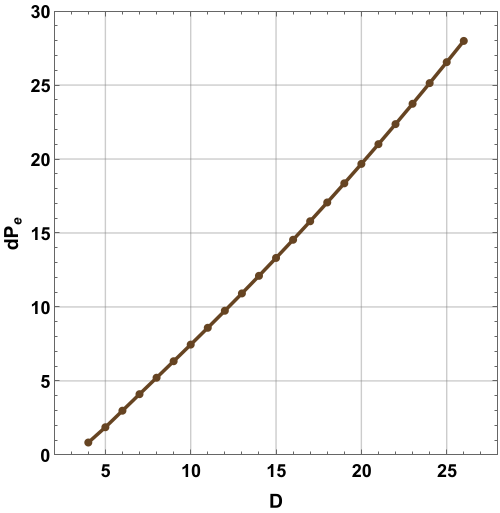}
		\caption{}
		\label{f5_2}	
	\end{subfigure}
	
	\caption{\footnotesize\it Entropy (a) and pressure (b) variation$dP$ as a function of spacetime dimension  with $\left| p^r_h\right| = 1$, $Q=1$, $l=1$ and $dQ = 0.1$. }
	\label{f5}
\end{figure}

In Fig.\ref{f5}, we depict the change of the differential entropy and pressure as a function of spacetime dimension for the extremal cases. 
We notice that differential entropy for the extremal case decreases for higher dimensional spacetime as it is shown in \eqref{40} while the variation of the pressure is always positive and gets larger for higher dimensional spacetime.

\section{The Weak cosmic censorship conjecture}

According to the WCCC, no singularities will be seen from asymptotic null infinity for the future observer. That is to say, a black hole event horizon must be required to shield singularities from an observer situated at spatial infinity. The weak cosmic censorship hypothesis must therefore be supported by the existence of an event horizon.
We will observe the black hole's ability to absorb a charged particle to determine if it has an event horizon. Specifically, we are going to discuss how $f(r)$ varies in such a scenario.

\begin{figure}[!ht]
	\centering \includegraphics[scale=0.7]{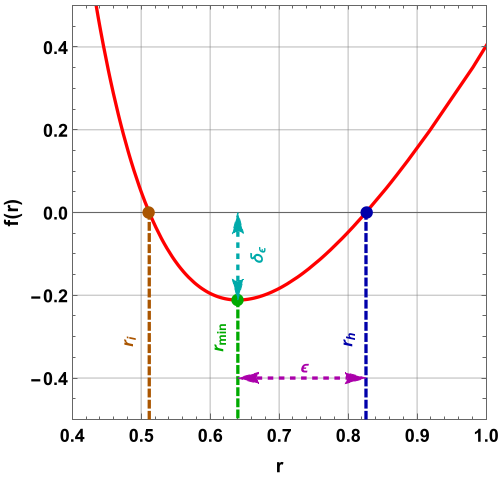}
	\caption{\footnotesize\it The variation of $f(r)$ function in terms of $r$ with $D= 4$, $Q=1$, $l=1$ and $M = 1.3$. }
	\label{f6}
\end{figure}

We illustrate in Fig.~\ref{f6} the behavior of the $f(r)$ function in terms of the radial coordinate $r$.

 There are two horizons, the inner (Cauchy) horizon located at $r_i$ and the event horizon (the outer horizon) at $r_h$. The metric function $f(r)$ decreases first for $r<r_{min}$, then admits a minimum at $r=r_{min}$. For $r>r_{min}$, the function is ever-increasing. At $r=r_{min}$, the function $f(r)$ must meet the following conditions
 \begin{equation}\label{43}
\begin{split}
&f(r_{min}) = f_{min} = \delta \le 0 \\
&\left. \dfrac{\partial f(r)}{\partial r}\right| _{r = r_{min}} = f'_{min} = 0\\
&\left. \dfrac{\partial^2 f(r)}{\partial r^2}\right| _{r = r_{min}} = f''_{min}  >0
\end{split}
\end{equation} 

We have a vanishing value $\delta = 0$ for the extreme black hole, i.e., when $r _i=r_h$, where $\delta$ is a sufficiently small quantity for near extremal configurations.
Both the Cauchy and event horizons remain on either side of $r_{min}$.
The associated black hole parameters, $M$, $Q$, and $l$ are transformed to $(M + dM; Q + dQ; l + dl)$ when the particle's charge is absorbed by the black hole.
As a result, the minimum value and event horizon will be located at $(r_{min} + dr_{min};r_h + dr_h)$.
Additionally, the transformation of $f(r)$ transcends to $df_{min}$.
Following the transformation, the new minimum point is encoded as
\begin{equation}\label{44}
\left. \dfrac{\partial f(r)}{\partial r}\right| _{r = r_{min} + dr_{min}} = f'_{min} + df'_{min} = 0,
\end{equation} 
which in turn gives
\begin{equation}\label{45}
df'_{min} = \dfrac{\partial f'_{min}}{\partial r_{min}} dr_{min} + \dfrac{\partial f'_{min}}{\partial M} dM + \dfrac{\partial f'_{min}}{\partial Q} dQ + \dfrac{\partial f'_{min}}{\partial l} dl = 0.
\end{equation}
The transformed relation for the function, $f( r_{min})$, reads 
\begin{equation}\label{46}
f( r_{min} + dr_{min}) = f_{min} + df_{min} = \delta + \left(  \dfrac{\partial f_{min}}{\partial M} dM + \dfrac{\partial f_{min}}{\partial Q} dQ + \dfrac{\partial f_{min}}{\partial l} dl \right). 
\end{equation}
We can impose the following conditions of near-extremal case such that
\begin{equation}\label{47}
\delta \to \delta_{\epsilon}, \quad r_h = r_{min} + \epsilon,
\end{equation}
that is to say that horizon of the near-extremal black hole is located slightly to the right of the minimum
point. It is to be mentioned that the minimum of the transformed function has a negative contribution and is arguably very small. This is given as $ \left| \delta_{\epsilon}\right| , \epsilon \ll 1$. For the near-extremal black hole, the transformed minimum value is
\begin{equation}\label{48}
f_{min}+df_{min}= \delta_{\epsilon} - \dfrac{2 \pi E r_{min}}{D-2} \left[ \dfrac{1-\Phi^2}{S} + \dfrac{4 r_{min}^3}{\left( D-3\right) V l^2}\right] + \mathcal{O}(\epsilon). 
\end{equation}
Forthwith, we focus on the extremal situation ($\delta_{\epsilon}, \epsilon =0$) and we remove $Q$ and $l$ terms by using $f_{min} = 0$. The transformed minimum value of the function $f(r)$ now becomes
\begin{equation}\label{49}
f_{min}+df_{min}= 0, \quad \delta_{\epsilon} = 0 \quad \epsilon = 0.
\end{equation}
Therefore, the extremal  black hole always stays at its stable minimum, which leads us to conclude that its phase cannot be changed, even if it is charged or discharged by the electrically charged particle absorption, as reported in \cite{art9}. 

\begin{figure}[!ht]
	\centering 
	\begin{subfigure}[h]{0.45\textwidth}
		\centering \includegraphics[scale=0.5]{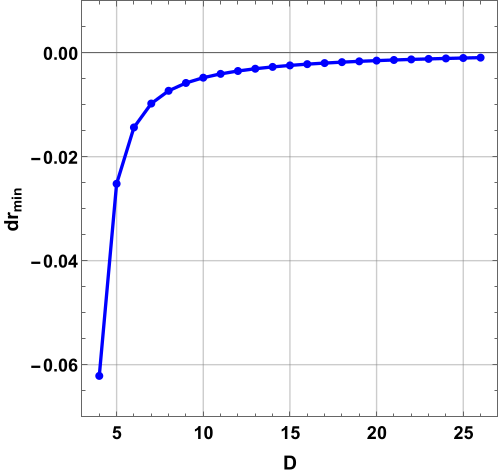}
		\caption{$dQ = 0.05$}
		\label{f7_1}
	\end{subfigure}
	\hspace{1pt}	
	\begin{subfigure}[h]{0.45\textwidth}
		\centering \includegraphics[scale=0.5]{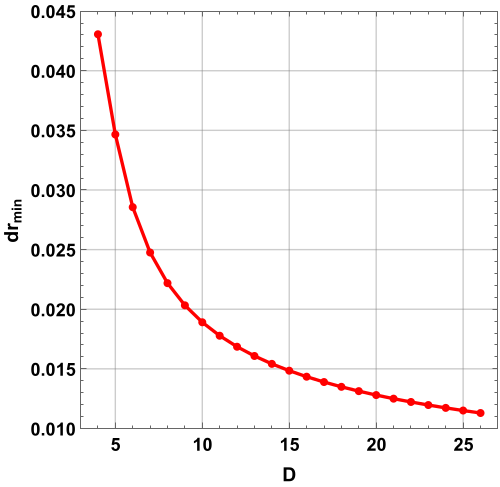}
		\caption{$dQ = 0.3$}
		\label{f7_2}	
	\end{subfigure}
	
	\caption{\footnotesize\it Variation of the minimum point $dr_{min}$ as a function of spacetime dimensions in extremal black holes for different absorbed particle charge $dQ$  with $\left| p^r_h\right| = 0.1$, $Q=1$ and $l=1$. }
	\label{f7}
\end{figure}

Although the above result seems to be correct apparently, there is a caveat in assuming that $dr_{min} = dr_h$. To see this discrepancy, we calculate $dr_{min}$ and compare it to $dr_h$. We compute the minimum point $ dr_{min}$ in extremal black holes using \eqref{32}, \eqref{39} and \eqref{45}, which reads
\begin{equation}\label{50}
dr_{min} = \dfrac{\left( D-2\right) \left( D-3\right) S r_h^3 \left[ E \Phi^2 +\left( D-2\right) \left(\left| p^r_h\right| - dQ \Phi\right) \right] -4 \pi \left( D-1\right)  E P V r_h^4 }{\left( D-2\right) \left( D-3\right) P V S \left[ \left( D-4\right) \left( D-1\right) r_h^2 +\left( D-3\right) l^2 \left( \left( 3 D-8\right) \Phi^2+\left( D-2\right) \right) \right]  },
\end{equation}
that can be put under the following form 
\begin{equation}\label{51}
dr_{min} = \dfrac{ r_h^3 \left[ E \Phi^2 +\left( D-2\right) \left(\left| p^r_h\right| - dQ \Phi\right) \right] -2  \left( D-1\right) P V r_h^2 dr_h }{ P V  \left[ \left( D-4\right) \left( D-1\right) r_h^2 +\left( D-3\right) l^2 \left( \left( 3 D-8\right) \Phi^2+\left( D-2\right) \right) \right]  },
\end{equation}
where we have used \eqref{40}. It is obvious from the above expression that $dr_{min}$ is different from $ dr_h$ and it could be negative also, particularly, when 
\begin{equation}\label{52}
dQ < \dfrac{\left| p^r_h\right|}{\Phi}.
\end{equation} 
The variation of the minimum point $dr_{min}$ as a function of spacetime dimensions for the extremal black holes for various values of the absorbed particle charge is depicted in  Fig.~\ref{f7}, 
and one can observe  from it that $dr_{min}$ is negative when $dQ = 0.05$ for all dimensions, so the minimum point is shifted to the left, whereas for $dQ = 0.3$, $dr_{min}$ is positive and the minimum point is displaced to the right. 

\begin{figure}[!ht]
	\centering 
	\begin{subfigure}[h]{0.45\textwidth}
		\centering \includegraphics[scale=0.5]{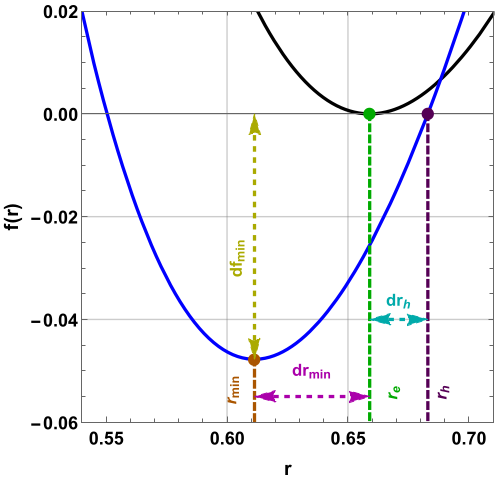}
		\caption{Shift to the left.}
		\label{f8_1}
	\end{subfigure}
	\hspace{1pt}	
	\begin{subfigure}[h]{0.45\textwidth}
		\centering \includegraphics[scale=0.5]{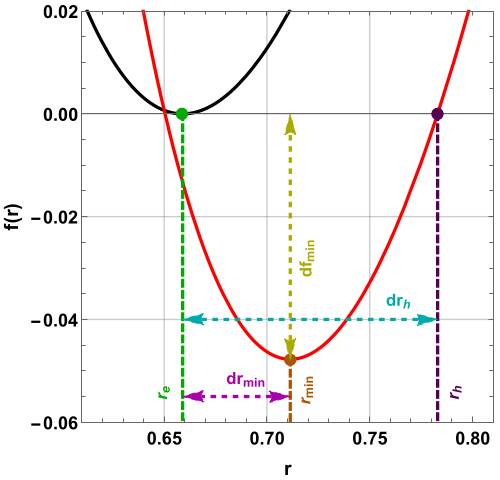}
		\caption{Shift to the right.}
		\label{f8_2}	
	\end{subfigure}
	
	\caption{\footnotesize\it Variation of $f(r)$ function after the particle absorption in the two situations: when the minimum point is shifted to the left and when it is shifted to the right. }
	\label{f8}
\end{figure}

Furthermore, we reveal in Fig.~\ref{f8}, the variation of $f(r)$ after the particle absorption process in the two different situations that we have introduced in the previous paragraph: $(a)$ when the minimum point is shifted to the left and $(b)$ when it is shifted to the right.

 It's remarked that the event horizon radius varies significantly and is always greater than the minimum point variation, $dr_h > dr_{min}$. When $dr_h = dr_{min}$, the extremal black hole still remains extremal even after the particle absorption. Therefore, for the extremal black hole, we can write 
\begin{equation}\label{53}
\begin{split}
df_{min} < 0 & \Longrightarrow dr_h > dr_{min},\\
df_{min} = 0 & \Longrightarrow dr_h = dr_{min},
\end{split}
\end{equation} 
Thus, if we want to look for the behavior of the function $f(r)$, it is sufficient to compare $dr_h$ and $dr_{min}$ for different parametric values. This in turn will lead us to know whether the black hole is still in the extremal or non-extremal state.

\begin{figure}[!ht]
	\centering 
	\begin{subfigure}[h]{0.45\textwidth}
		\centering \includegraphics[scale=0.5]{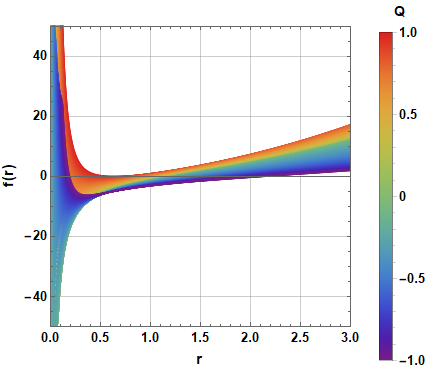}
		\caption{}
		\label{f9_1}
	\end{subfigure}
	\hspace{1pt}	
	\begin{subfigure}[h]{0.45\textwidth}
		\centering \includegraphics[scale=0.5]{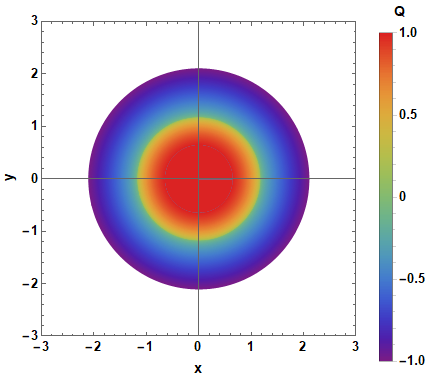}
		\caption{}
		\label{f9_2}	
	\end{subfigure}
	\begin{subfigure}[h]{0.45\textwidth}
	\centering \includegraphics[scale=0.5]{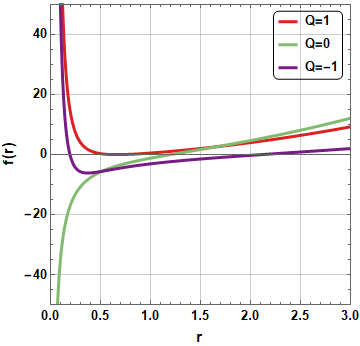}
	\caption{}
	\label{f9_3}
\end{subfigure}
\hspace{1pt}	
\begin{subfigure}[h]{0.45\textwidth}
	\centering \includegraphics[scale=0.5]{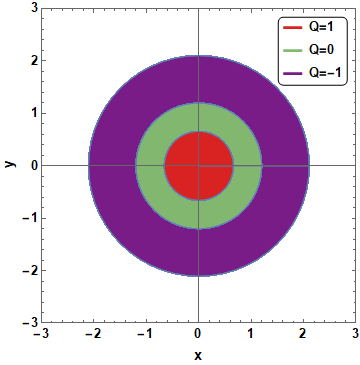}
	\caption{}
	\label{f9_4}	
\end{subfigure}
	
	\caption{\footnotesize\it (a) and (c): Evolution of $f(r)$ function during absorption of $10^3$ particles with a fixed electric charge $dQ = -0.002$. (b) and (d): Evolution of black hole projection on $xy$ plan during absorption of $10^3$ particles with a fixed electric charge $dQ = -0.002$. The initial conditions are $Q=1$, $l=1$, $M = 1.23$, $r_h= 0.65$ and $D=4$. }
	\label{f9}
\end{figure}

Next, we intend to study numerically a neutralization and recharging of an extremal black hole. We begin with an extremal black hole with $Q=1$ which will absorb an estimated $10^3$ negatively charged particles with constant electric charge $dQ = -0.002$. It is clear from the radial momentum that it should depend on the event horizon radius such that 
\begin{equation}\label{54}
\left| p^r_h\right| = 2 \left| dQ \Phi \right|, 
\end{equation} 
besides, in order to ensure that the particle energy is positive. We illustrate in Fig.~\ref{f9} the evolution of  $f(r)$ function and the black hole projection on $x-y$ plane during the absorption of $10^3$ particles where every particle has a negative charge $dQ = -0.002$.
In the beginning, we have an extremal black hole with $Q=1$, the black hole eventually absorbs negative charge and breaks down to a non-extremal state, and continues absorbing particles till it becomes Schwarzschild-like after it has absorbed $500$ particles. Moreover, the black hole charge becomes negative till it flips the sign of the initial value and finally the charge becomes $Q = -1$.

\begin{figure}[!ht]
	\centering 
	\begin{subfigure}[h]{0.45\textwidth}
		\centering \includegraphics[scale=0.5]{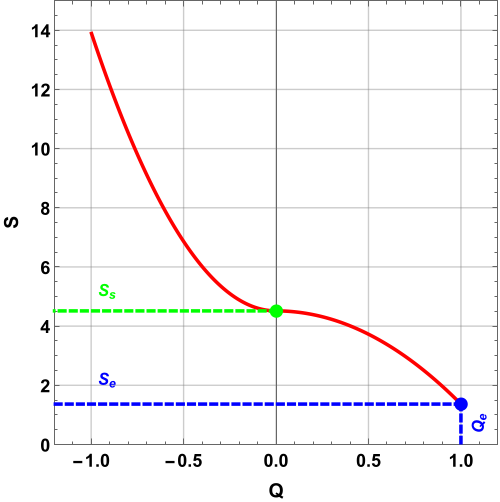}
		\caption{Black hole entropy as a function of electric charge.}
		\label{f10_1}
	\end{subfigure}
	\hspace{1pt}	
	\begin{subfigure}[h]{0.45\textwidth}
		\centering \includegraphics[scale=0.5]{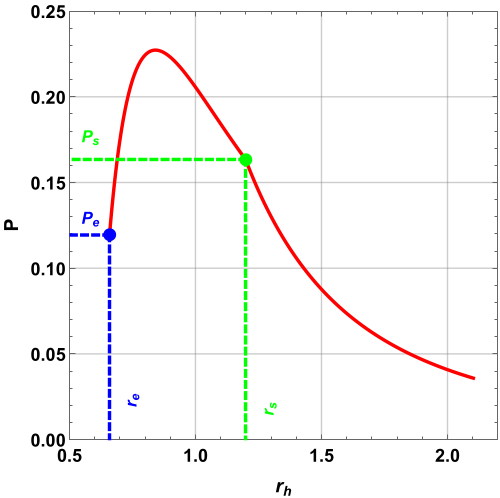}
		\caption{Black hole pressure as a function of the black hole event horizon.}
		\label{f10_2}	
	\end{subfigure}
	\begin{subfigure}[h]{0.45\textwidth}
		\centering \includegraphics[scale=0.5]{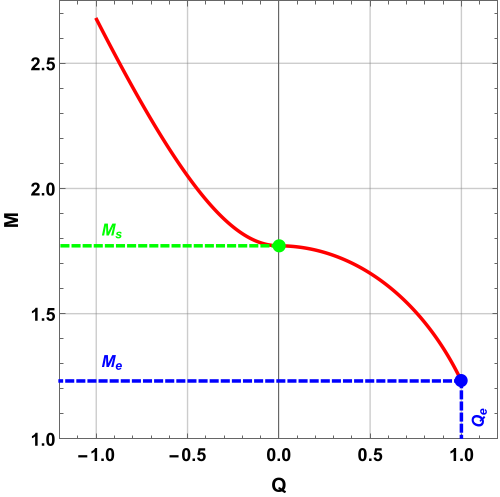}
		\caption{Black hole mass as a function of electric charge.}
		\label{f10_3}
	\end{subfigure}
	\hspace{1pt}	
	\begin{subfigure}[h]{0.45\textwidth}
		\centering \includegraphics[scale=0.5]{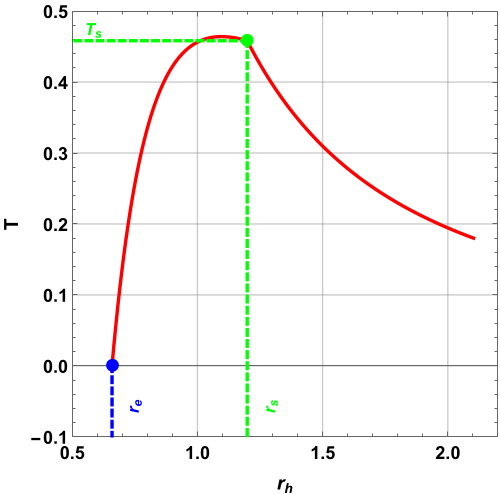}
		\caption{Black hole temperature as a function of the black hole event horizon. }
		\label{f10_4}	
	\end{subfigure}
	
	\caption{\footnotesize\it Evolution of different thermodynamical variables during an absorption of $10^3$ particles with a fixed electric charge $dQ = -0.002$. The initial conditions are $Q=1$, $l=1$, $M = 1.23$, $r_h= 0.65$ and $D=4$. }
	\label{f10}
\end{figure}

Proceeding with similar arguments, we show in Fig.\ref{f10} the evolution of different thermodynamic variables during the absorption of $10^3$ particles with a fixed electric charge $dQ = -0.002$.
 We have begun with an extremal black hole in a four-dimensional spacetime with $Q=1$ and $l=1$. We observe that the entropy increases during the absorption of particles and thereby verifying the second law which is found to held. Moreover, the entropy has an inflection point where the black hole gets neutralized. The black hole's pressure increases near the extremal case as it is expected, and then it begins to decrease and changes its concavity near $r_h=r_s = 1.2$ corresponding to the neutralized black hole case. In addition, the inflection point in the pressure curve corresponds to a discontinuity of its derivative that indicates a possible phase transition at this point. Subsequent analysis of the black hole mass and entropy evolution share the same qualitative behavior. Whereas, the black hole temperature profile shows a clear phase transition at $r_h=r_s$. Indeed, for $r_h<r_s$ the black hole is positively charged and hence absorbs negative charges to be neutralized. We see that the temperature increases  before the phase transition, but when the black hole is neutralized ($r_h=r_s$), the temperature switches off both its monotony and concavity indicating a discontinuity in its first derivative, and therefore represents a hallmark of phase transition. 
 
\begin{figure}[h!]
	\centering 
	\begin{subfigure}[h]{0.45\textwidth}
		\centering \includegraphics[scale=0.5]{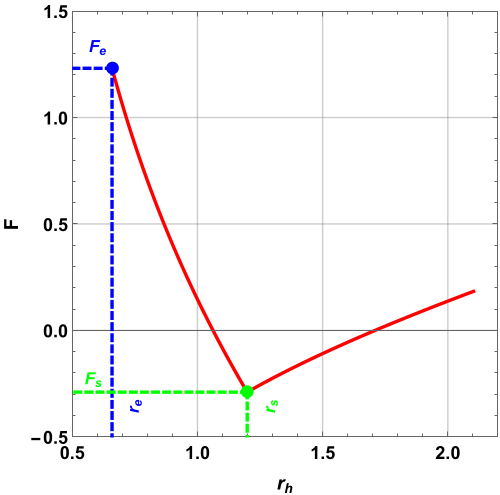}
		\caption{}
		\label{f11_1}
	\end{subfigure}
	\hspace{1pt}	
	\begin{subfigure}[h]{0.45\textwidth}
		\centering \includegraphics[scale=0.5]{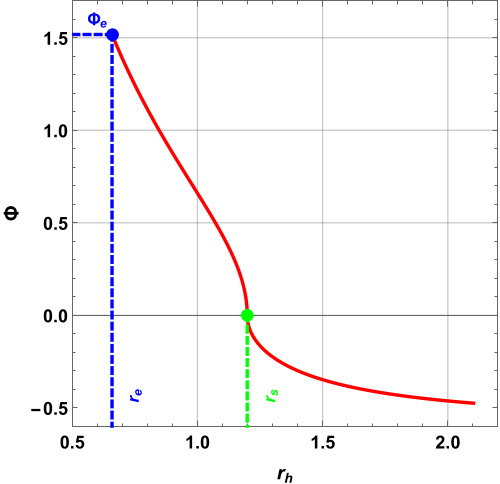}
		\caption{}
		\label{f11_2}	
	\end{subfigure}
	
	\caption{\footnotesize\it (a) Free energy and (b) electric potential as a function of black hole event horizon during absorption of $10^3$ particles with a fixed electric charge $dQ = -0.002$. The initial conditions are $Q=1$, $l=1$, $M = 1.23$, $r_h= 0.65$ and $D=4$. }
	\label{f11}
\end{figure}

 Furthermore, to consolidate with such a phase transition behavior we recall the free energy portrait as a function of the black hole event horizon as is shown in Fig.\ref{f11}.
 Clearly, the $F-r_h$ diagram unveils a discontinuity in its derivative  affirming that a phase transition occurs when the black hole changes the sign of its charge. Additionally, we notice that the electric potential has an inflection point where it vanishes. Therefore, the electric potential changes its sign at this point, where it alters its concavity too.

\section{Black hole neutralization}

In this section, we attend to examine the black hole neutralization, concretely,  the neutralization of an extremal black hole and the fluctuations around the Schwarzchild-like state (vanishing charge solution) following a variety of thermodynamics processes.

\subsection{Isentropic neutralization}

Herein, we discuss the isentropic process where the entropy of any thermodynamic system remains constant. Such a process happens when the thermodynamic system is both adiabatic and reversible. Using Eq.\eqref{40} and Eq.\eqref{19}, we have an isentropic neutralization when the electric charge $q = dQ$ of the absorbed particle has the value\footnote{Which implies that the charge of the absorbed particle has the inverse sign of the black hole charge, and thereby leading to the neutralization of the black holes and the electrically charged particle system.} 
 \begin{equation}\label{55}
 q = -  \dfrac{ \left| p^r_h\right| r_h^{D-3} }{Q_D},
\end{equation}

\begin{figure}[!ht]
	\centering 
	\begin{subfigure}[h]{0.45\textwidth}
		\centering \includegraphics[scale=0.5]{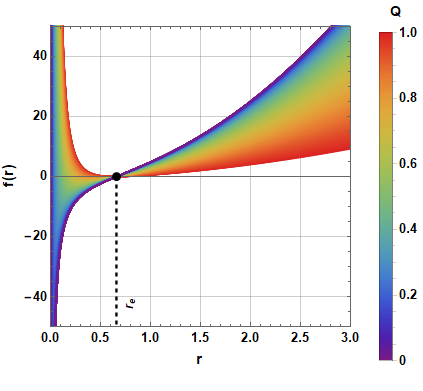}
		\caption{}
		\label{f12_1}
	\end{subfigure}
	\hspace{1pt}	
	\begin{subfigure}[h]{0.45\textwidth}
		\centering \includegraphics[scale=0.5]{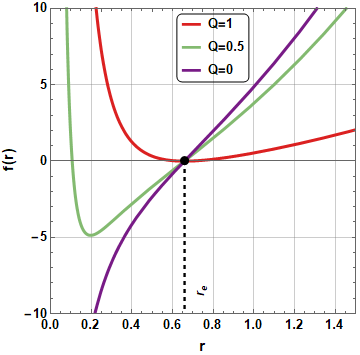}
		\caption{}
		\label{f12_2}	
	\end{subfigure}

	\caption{\footnotesize\it Evolution of $f(r)$ function during an isentropic absorption of $500$ particles with a fixed electric charge $dQ = -0.002$. The initial conditions are $Q=1$, $l=1$, $M = 1.23$, $r_h= 0.65$ and $D=4$. }
	\label{f12}
\end{figure}

\begin{figure}[!ht]
	\centering 
	\begin{subfigure}[h]{0.45\textwidth}
		\centering \includegraphics[scale=0.5]{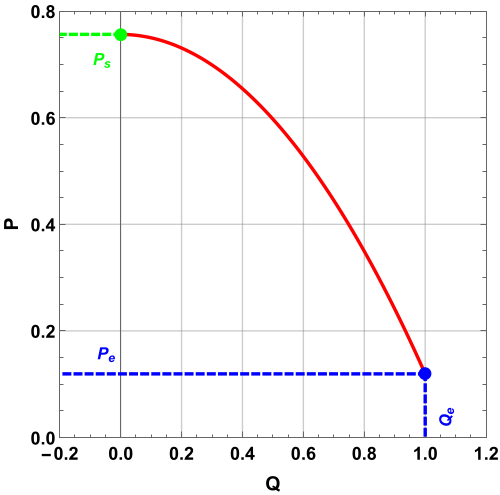}
		\caption{Pressure as a function of black hole electric charge.}
		\label{f13_1}
	\end{subfigure}
	\hspace{1pt}	
	\begin{subfigure}[h]{0.45\textwidth}
		\centering \includegraphics[scale=0.5]{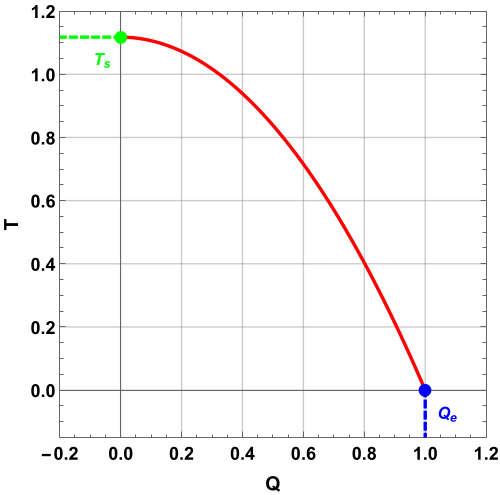}
		\caption{Temperature as a function of black hole electric charge.}
		\label{f13_2}	
	\end{subfigure}
	\begin{subfigure}[h]{0.45\textwidth}
		\centering \includegraphics[scale=0.5]{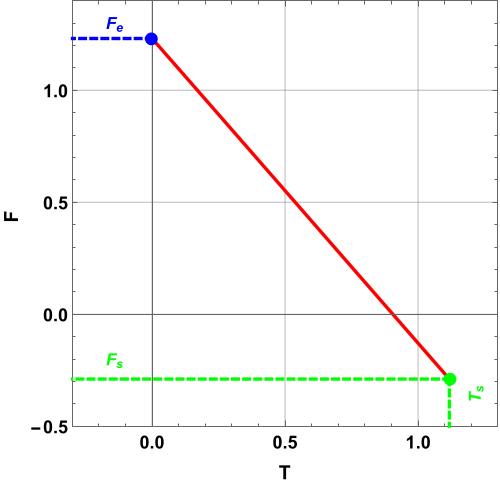}
		\caption{Free energy as a function of black hole electric charge.}
		\label{f13_3}
	\end{subfigure}
	\hspace{1pt}	
	\begin{subfigure}[h]{0.45\textwidth}
		\centering \includegraphics[scale=0.5]{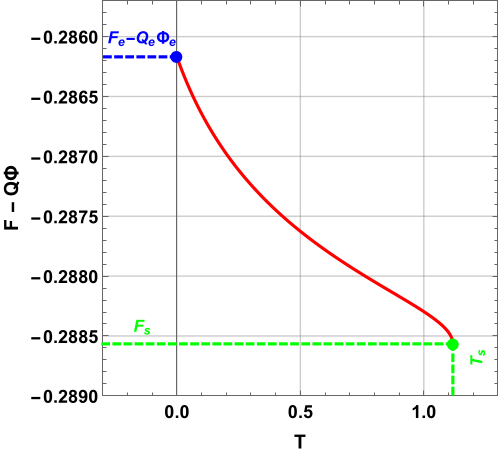}
		\caption{Grand potential as a function of black hole electric charge. }
		\label{f13_4}	
	\end{subfigure}
	
	\caption{\footnotesize\it Evolution of different thermodynamical variables during an isentropic absorption of $500$ particles with a fixed electric charge $dQ = -0.002$. The initial conditions are $Q=1$, $l=1$, $M = 1.23$, $r_h= 0.65$ and $D=4$. }
	\label{f13}
\end{figure}

We begin to discuss by looking at the isentropic relaxation to the neutral state. We depict in Fig.~\ref{f12}, the neutralization of an extremal black hole in four AdS spacetime dimensions which absorbs $500$ particles with $dQ = - 0.002$ respecting Eq.\eqref{55}. 
When the extremal black hole begins to be neutralized, it breaks down to a non-extremal state keeping the size of the event horizon radius the same, but the inner horizon radius decreases. Moreover, the mass and the entropy are constant.  Next, we illustrate in Fig.\ref{f13} different thermodynamic quantities characterizing the isentropic neutralization of an extremal black hole.

The $(a)$ and $(b)$ panels reveal that the pressure and temperature of the black hole share qualitatively the same behavior, indeed they increase during the neutralization and reach their maximum value, $P_s$ and $T_s$ when the black hole is completely neutralized. Moreover, the temperature becomes non-null once the black hole has absorbed a charge and it breaks into a non-extremal black hole. From the $(c)$ panel, we notice that the free energy decreases linearly in terms of temperature as it is expected while keeping the mass and the entropy constant. The potential in the grand canonical ensemble, $F-Q\Phi$, decreases quicker to reach the critical point where a phase transition is expected as shown in the $(d)$ panel.

\begin{figure}[!ht]
	\centering 
	\begin{subfigure}[h]{0.45\textwidth}
		\centering \includegraphics[scale=0.4]{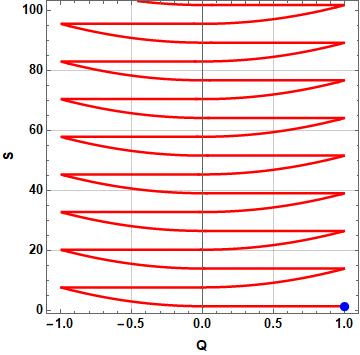}
		\caption{Entropy as a function of black hole electric charge.}
		\label{f14_1}
	\end{subfigure}
	\hspace{1pt}	
	\begin{subfigure}[h]{0.45\textwidth}
		\centering \includegraphics[scale=0.4]{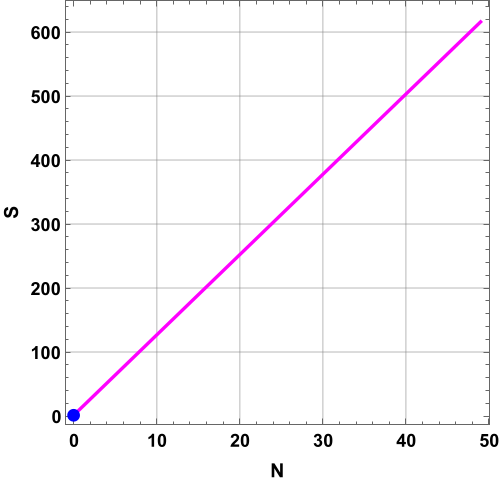}
		\caption{Entropy as a function of the number of periods.}
		\label{f14_2}	
	\end{subfigure}
	\begin{subfigure}[h]{0.45\textwidth}
		\centering \includegraphics[scale=0.4]{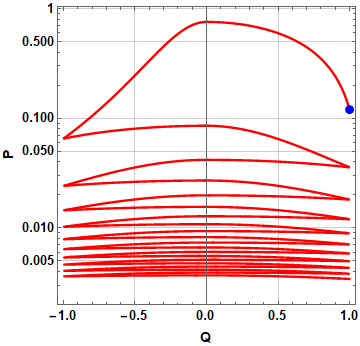}
		\caption{Pressure as a function of black hole electric charge.}
		\label{f14_3}
	\end{subfigure}
	\hspace{1pt}	
	\begin{subfigure}[h]{0.45\textwidth}
		\centering \includegraphics[scale=0.4]{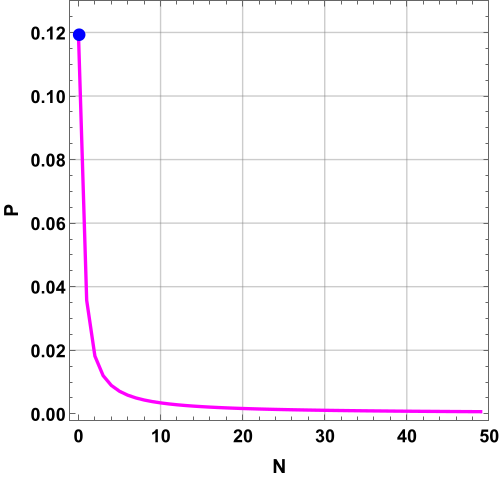}
		\caption{Pressure as a function of the number of periods. }
		\label{f14_4}	
	\end{subfigure}
	\begin{subfigure}[h]{0.45\textwidth}
	\centering \includegraphics[scale=0.4]{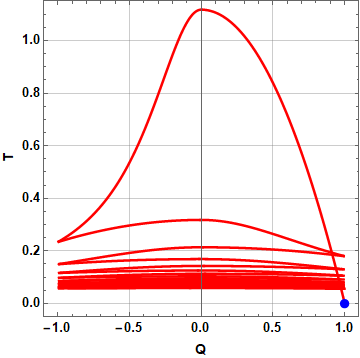}
	\caption{Temperature as a function of black hole electric charge.}
	\label{f14_5}
\end{subfigure}
\hspace{1pt}	
\begin{subfigure}[h]{0.45\textwidth}
	\centering \includegraphics[scale=0.45]{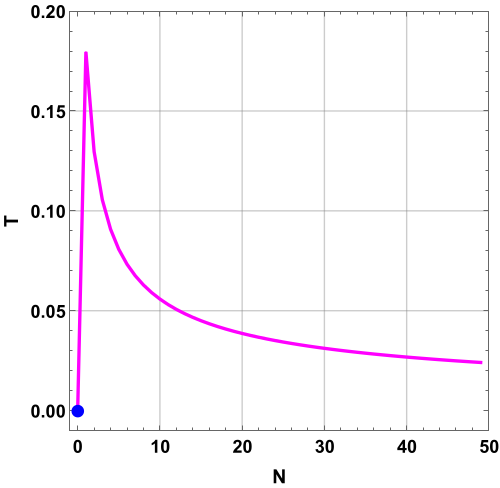}
	\caption{Temperature as a function of the number of periods. }
	\label{f14_6}	
\end{subfigure}
	
	\caption{\footnotesize\it Evolution of different thermodynamical variables during absorption of $10^5$ particles over $50$ periods. The initial conditions (blue dot) are $Q=1$, $l=1$, $M = 1.23$, $r_h= 0.65$, and $D=4$. }
	\label{f14}
\end{figure}

The next relevant step is to look at the oscillations of the black hole around the Scwarzchild-like state with constant amplitude. This means that we begin with an extremal black hole state in which the value of the electric charge varies between $+Q$ and $-Q$. We simulate the absorption of $10^5$ particles over $N=50$ periods. Within each period, the black hole gets neutralized through the isentropic process, then gains its charge and ultimately it becomes charged with the opposite signature but with the same magnitude, after, it gets neutralized isentropically again and recharged with the same charge as it was at the beginning. The results of the simulation are displayed in Fig.\ref{f14}.

\begin{figure}[!ht]
	\centering 
	\begin{subfigure}[h]{0.45\textwidth}
		\centering \includegraphics[scale=0.5]{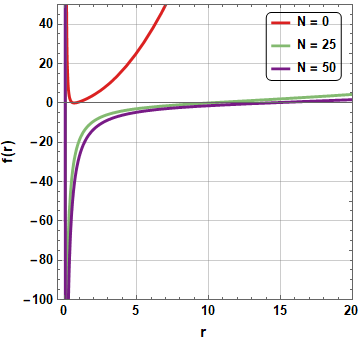}
		\caption{}
		\label{f15_3}
	\end{subfigure}
	\hspace{1pt}	
	\begin{subfigure}[h]{0.45\textwidth}
		\centering \includegraphics[scale=0.5]{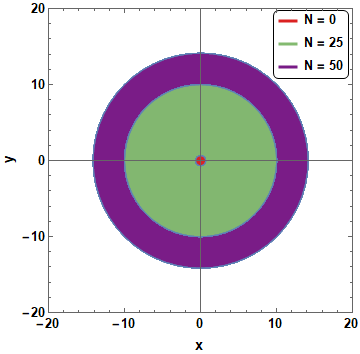}
		\caption{}
		\label{f15_4}	
	\end{subfigure}
	\caption{\footnotesize\it (a) : Evolution of $f(r)$ function during an absorption of $10^5$ over $50$ periods . (b) : Evolution of black hole projection on $xy$ plan during an absorption of $10^5$ over $50$ periods. The initial conditions are $Q=1$, $l=1$, $M = 1.23$, $r_h= 0.65$ and $D=4$. }
	\label{f15}
\end{figure}

We see that the entropy increases at each period with a fixed quantity, $\Delta S = 12.55$, so it increases linearly in terms of the number of periods, which means that the black hole gets larger. While the pressure increases within every neutralization but it decreases quickly when the black hole gets charged. Hence, it decreases quickly in terms of the number of periods and tends to zero which means that the geometry of spacetime tends to be flat. Furthermore, the vanishing temperature associated with the extremal case at the beginning of the simulation increases quickly during the first period where we observe a peak that corresponds to the breaking of the extremal black to a non-extremal one. Then it decreases while the black hole gets larger in size. We mention that the increase of the temperature does not mean that the black hole tends to be extremal, because when the black hole gets larger, it goes so far from the extremal state as illustrated in Fig.\ref{f15},
 where the extremal black hole (red curve) becomes very large after $25$ period (green curve). Moreover, when the event horizon radius gets larger, the inner horizon radius gets smaller such as after $50$ period (purple curve) the event horizon is so far from the inner horizon that tends to zero.

To probe the phase transitions, we plot in Fig.~\ref{16} the behavior of different thermodynamic potentials during the absorption of $10^5$ particles over $50$ periods.
\begin{figure}[!ht]
	\centering 
	\begin{subfigure}[h]{0.45\textwidth}
		\centering \includegraphics[scale=0.42]{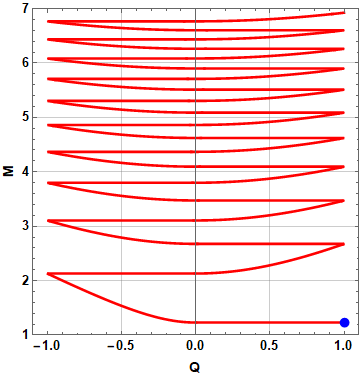}
		\caption{Mass as a function of black hole electric charge.}
		\label{f16_1}
	\end{subfigure}
	\hspace{1pt}	
	\begin{subfigure}[h]{0.45\textwidth}
		\centering \includegraphics[scale=0.42]{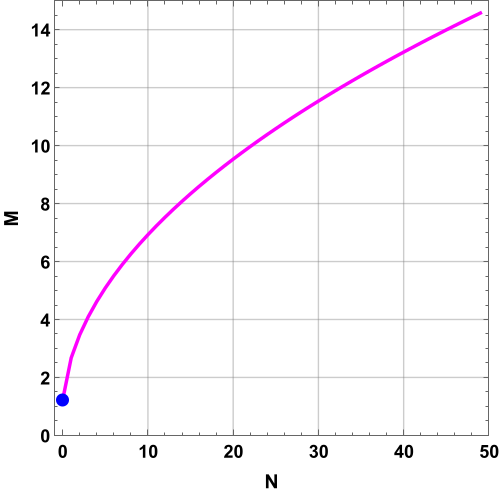}
		\caption{Mass as a function of the number of periods.}
		\label{f16_2}	
	\end{subfigure}
	\begin{subfigure}[h]{0.45\textwidth}
		\centering \includegraphics[scale=0.42]{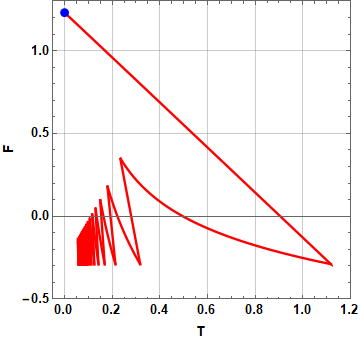}
		\caption{Free energy as a function of black hole electric charge.}
		\label{f16_3}
	\end{subfigure}
	\hspace{1pt}	
	\begin{subfigure}[h]{0.45\textwidth}
		\centering \includegraphics[scale=0.42]{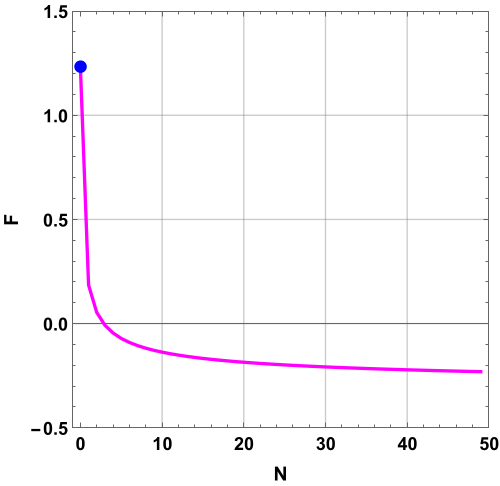}
		\caption{Free energy as a function of the number of periods. }
		\label{f16_4}	
	\end{subfigure}
	\begin{subfigure}[h]{0.45\textwidth}
		\centering \includegraphics[scale=0.42]{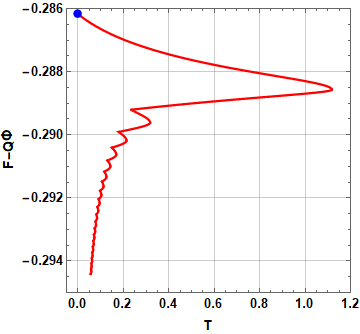}
		\caption{Grand potential as a function of black hole electric charge.}
		\label{f16_5}
	\end{subfigure}
	\hspace{1pt}	
	\begin{subfigure}[h]{0.45\textwidth}
		\centering \includegraphics[scale=0.42]{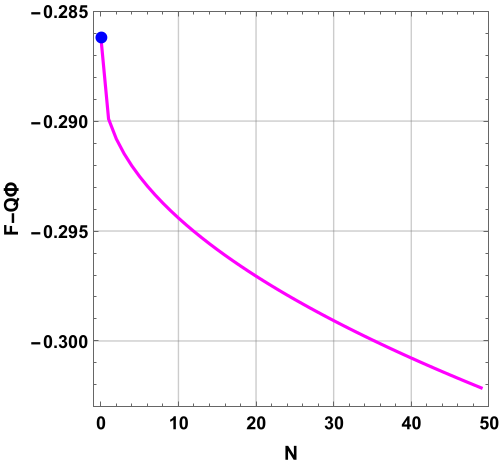}
		\caption{Grand as a function of the number of periods. }
		\label{f16_6}	
	\end{subfigure}
	
	\caption{\footnotesize\it Evolution of different thermodynamic potentials during an absorption of $10^5$ particles over $50$ periods. The initial conditions (blue dot) are $Q=1$, $l=1$, $M = 1.23$, $r_h= 0.65$, and $D=4$. }
	\label{f16}
\end{figure}
 One observes that the mass is constant during every process of neutralization but it increases when the black hole gets charged. The quantity $\Delta M$ reflects that the mass increases at each period is not constant and decreases with the number of the periods grows. 

 Nevertheless, the mass becomes more important when the black hole becomes larger. The free energy and the grand canonical potential curves decrease as function temperature when the black hole breaks into a non-extremal state and a phase transition occurs when the black hole is neutralized. Such behavior is repeated qualitatively in each period and it is amortized as the black hole gets larger, i.e., the electric charge becomes negligible compared to the mass. Both free energy and grand potential decrease quickly in terms of the number of periods, which indicates that the black hole becomes more stable.

\begin{figure}[!ht]
	\centering 
	\begin{subfigure}[h]{0.45\textwidth}
		\centering \includegraphics[scale=0.5]{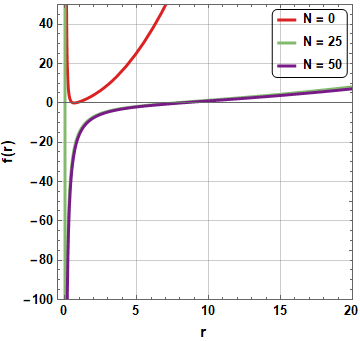}
		\caption{}
		\label{f17_1}
	\end{subfigure}
	\hspace{1pt}	
	\begin{subfigure}[h]{0.45\textwidth}
		\centering \includegraphics[scale=0.5]{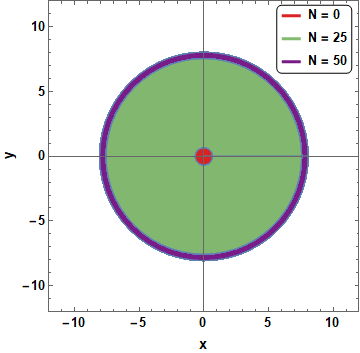}
		\caption{}
		\label{f17_2}	
	\end{subfigure}
	
	\caption{\footnotesize\it (a): Evolution of $f(r)$ function during an isentropic neutralization of an extremal black hole with the absorption of $10^5$ over $50$ periods. (b) : Evolution of black hole projection on $xy$ plan during an isentropic neutralization of an extremal black hole with an absorption of $10^5$ over $50$ periods. The initial conditions are $Q=1$, $l=1$, $M = 1.23$, $r_h= 0.65$ and $D=4$. }
	\label{f17}
\end{figure}

\begin{figure}[H]
	\centering 
	\begin{subfigure}[h]{0.45\textwidth}
		\centering \includegraphics[scale=0.4]{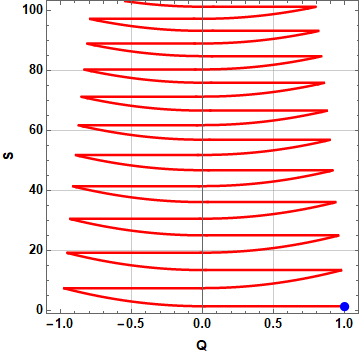}
		\caption{Entropy as a function of black hole electric charge.}
		\label{f18_1}
	\end{subfigure}
	\hspace{1pt}	
	\begin{subfigure}[h]{0.45\textwidth}
		\centering \includegraphics[scale=0.4]{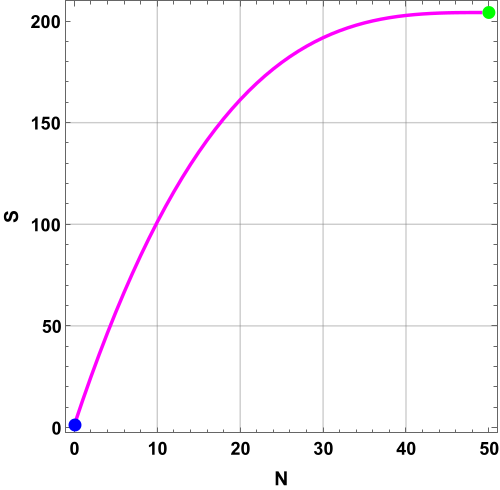}
		\caption{Entropy as a function of the number of periods.}
		\label{f18_2}	
	\end{subfigure}
	\begin{subfigure}[h]{0.45\textwidth}
		\centering \includegraphics[scale=0.4]{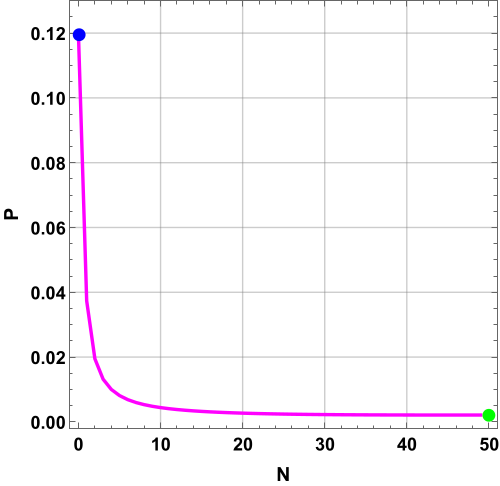}
		\caption{Pressure as a function of the number of periods.}
		\label{f18_3}
	\end{subfigure}
	\hspace{1pt}	
	\begin{subfigure}[h]{0.45\textwidth}
		\centering \includegraphics[scale=0.4]{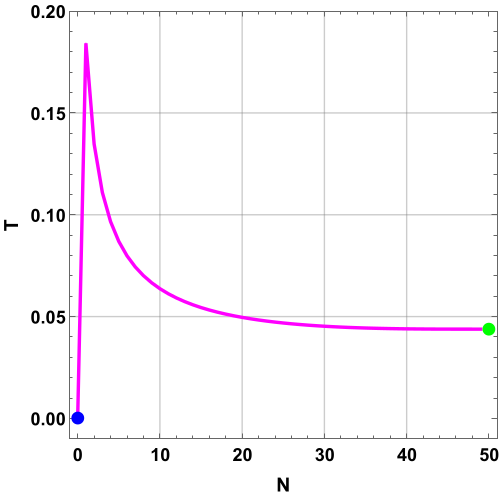}
		\caption{Temperature as a function of the number of periods. }
		\label{f18_4}	
	\end{subfigure}
	\begin{subfigure}[h]{0.45\textwidth}
		\centering \includegraphics[scale=0.4]{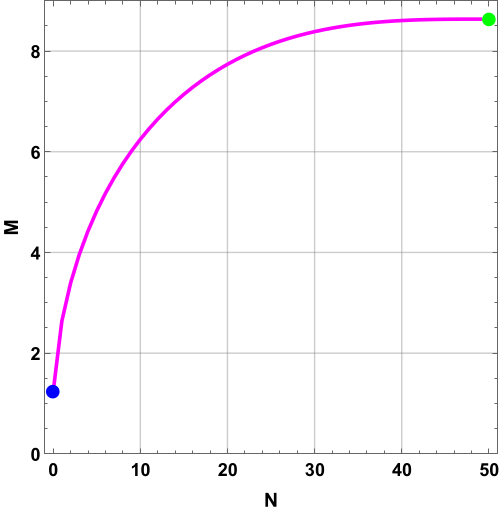}
		\caption{Mass as a function of the number of periods.}
		\label{f18_5}
	\end{subfigure}
	\hspace{1pt}	
	\begin{subfigure}[h]{0.45\textwidth}
		\centering \includegraphics[scale=0.4]{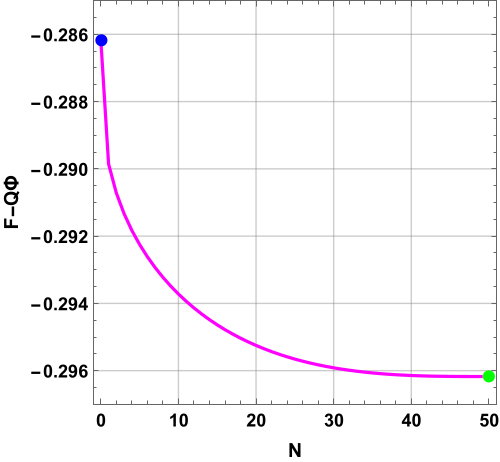}
		\caption{Grand potential as a function of the number of periods. }
		\label{f18_6}	
	\end{subfigure}
	
	\caption{\footnotesize\it Evolution of different thermodynamical quantity during an isentropic neutralization with the absorption of $10^5$ particles over $50$ periods. The final state which is an AdS-Schwarzchild black hole is represented by a green dot.  The initial conditions (blue dot) are $Q=1$, $l=1$, $M = 1.23$, $r_h= 0.65$, and $D=4$. }
	\label{f18}
\end{figure}

Let us now consider the neutralization of an extremal black hole using the same process but the amplitude of the oscillations decreases with the number of periods to get a Schawrzchild-like state at the end. That to say, we begin with a charge of $Q = +Q_e$, the black hole will be neutralized through the isentropic process and gets charged negatively with $Q = - (N-1) Q_e/N $, then it will be neutralized again through the same isentropic process, and gets charged positively with $Q = + (N-1) Q_e/N $, where $N$ is the number of periods used in the simulation. We repeat this process each period ($N$ periods) until we get $Q = 0$ and the black hole stops oscillating. The Fig.\ref{f17} shows the evolution of the function $f(r)$ and the projection of the black hole on $x-y$ plane during an isentropic neutralization of an extremal black hole with an absorption of $10^5$ over $50$ periods in the usual four dimensions.
The black hole size grows quickly before forming a Schwarzchild-AdS black hole. Finally, the grand potential decreases very quickly to reach its minimum value which corresponds to the stable point and the preferred state.  Afterward, we examine the evolution of different thermodynamic quantities that are displayed in Fig.~\ref{f18}.

The figure reveals that the entropy loses its previous linear behavior as is seen in the case of free oscillations and now it increases monotonously respecting the second law and converges to the stable point (green dot) which is the Schwarzchild-AdS state. The pressure decreases very quickly and which means that the AdS radius gets very large. In addition, the temperature unveils qualitatively the same behavior as in free oscillations. Indeed, it increases when the black hole becomes non-extremal and decreases to reach its minimum at the Schwarzchild-AdS state. Besides, the mass increases quickly in the beginning because the black absorbs the kinetic and the electromagnetic energy of the particle, but when the black approach the Schwarzschild-AdS state, the mass converges to its maximum value because the absorbed electromagnetic energy becomes very small and the absorbed kinetic energy as well.

Before ending this subsection, we should mention that the isentropic neutralization is an adiabatic and reversible process, such an idealization does not possibly occurs in the case of a black hole. Absorption of a particle is an irreversible process and the variation of the entropy should be strictly positive $dS>0$. That is the proposal that we will attend to discuss in the next subsection.

\subsection{Effective entropy}

In this subsection, we turn our attention to the effective thermodynamics that has attracted increasing interest, particularly in the context of dS black holes and thermodynamics of multiple horizons \cite{art2, Chabab:2020xwr, art17, art18, art19, art20}. 
In the case of multiple horizons context, all of the horizons should be
 taken into account and treated simultaneously as the first laws of black holes should be held equally at each horizon. In what follows, we will recall the effective approach to reveal that effective entropy varies when we have an isentropic neutralization.  The absorption of a particle that respects the relation Eq.\eqref{55} is a reversible process since the entropy is still constant in such cases.

In our considered black hole model, two horizons persist, namely, the inner horizon $r_i$ and the vent horizon (outer horizon). The first law of thermodynamics associated with the event horizon is given in Eq.\eqref{9} but for the inner horizon, its first law reads
\begin{equation}\label{56}
dM = -T_i dS_i+V_i dP + \Phi_i dQ ,
\end{equation}
where the different thermodynamical variables that describe the inner horizon are given by 
\begin{equation}\label{57}
T_i = - \dfrac{1}{4 \pi } \left( \dfrac{\partial f(r)}{\partial r}\right)_{r=r_i}   , \quad S = \dfrac{\Omega_{D-2}r_i^{D-2}}{4} \quad \Phi_i = \dfrac{Q_D}{r_i^{D-3}} \quad \text{and} \quad V_i = \dfrac{\Omega_{D-2}}{D-1}r_i^{D-1}.
\end{equation}
The variation of the black hole mass during an absorption of a charged particle can be written
\begin{equation}\label{58}
dM = \dfrac{\partial M}{\partial r_i} dr_i + \dfrac{\partial M}{\partial Q} dQ + \dfrac{\partial M}{\partial l} dl.
\end{equation}
Using Eq.\eqref{32} and Eq.\eqref{39}, we can express $dr_i$ in terms of $r_i$, $r_h$, $Q$, $l$, $dQ$ and $D$, thereby the variation of the inner entropy, $dS_i$ \footnote{We do not write down the expression of $dr_i$ and $dS_i$ because they are large and they are not necessary.}.

\begin{figure}[!ht]
	\centering 
	\begin{subfigure}[h]{0.45\textwidth}
		\centering \includegraphics[scale=0.45]{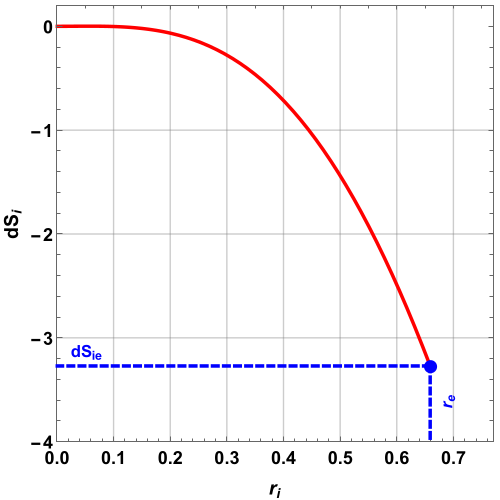}
		\caption{$D=4$}
		\label{f19_1}
	\end{subfigure}
	\hspace{1pt}	
	\begin{subfigure}[h]{0.45\textwidth}
		\centering \includegraphics[scale=0.45]{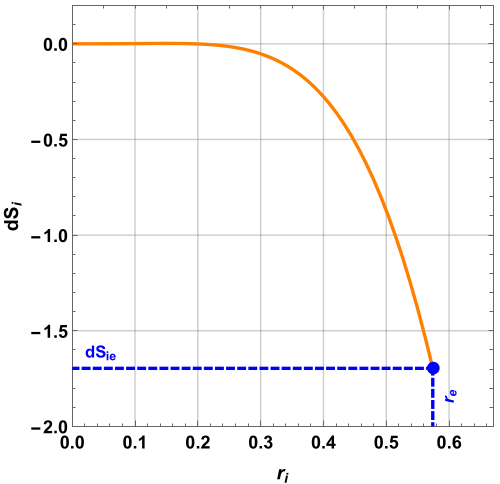}
		\caption{$D=5$}
		\label{f19_2}	
	\end{subfigure}
	\hspace{1pt}	
	\begin{subfigure}[h]{0.45\textwidth}
		\centering \includegraphics[scale=0.45]{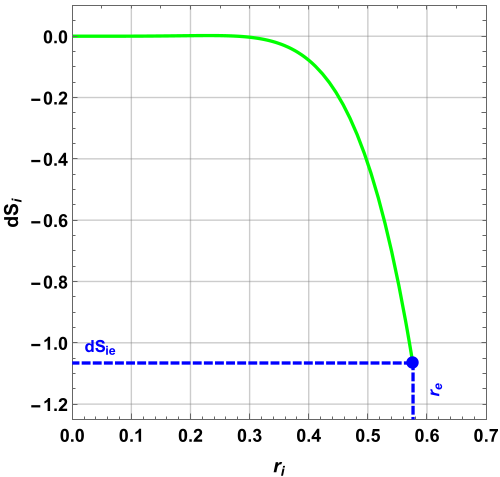}
		\caption{$D=6$}
		\label{f19_3}	
	\end{subfigure}
	\hspace{1pt}	
	\begin{subfigure}[h]{0.45\textwidth}
		\centering \includegraphics[scale=0.45]{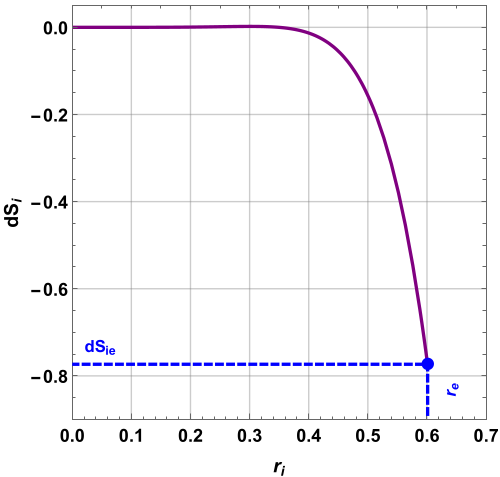}
		\caption{$D=7$}
		\label{f19_4}	
	\end{subfigure}
	
	\caption{\footnotesize\it Inner entropy variation $dS_i$ as a function of inner horizon radius $r_i$ for different dimensions  with $\left| p^r_h\right| = 1$, $Q=1$, $l=1$ and $dQ = 0.1$. The event horizon is fixed to the extremal radius. }
	\label{f19}
\end{figure}

In Fig.\ref{f19}, we depict the inner entropy variation $dS_i$ as a function of inner horizon radius $r_i$ for different dimensions where the event horizon is fixed to the extremal radius which corresponds to an isentropic process (the entropy associated with the event is fixed). 

From this figure, one can notice that the differential change in the inner entropy is ever decreasing and negative for all dimensions and decreases when the inner horizon increases. The minimum value corresponds to the extremal state. However, the differential change in the inner entropy vanishes when the inner horizon tends to zero which corresponds to a Schwarzchild-like state \footnote{Schwarzchild-like state does not have an inner horizon, but the null inner horizon radius corresponds to a Schwarzchild-like black hole}. The fact that the infinitesimal change in the inner entropy is negative which does not guarantee the violation of the second law, but it confirms it, such confirmation comes from the minus sign in front of the entropy in the first law of thermodynamics given in Eq.\eqref{56}.

Through the effective thermodynamics framework, the thermodynamics first law \cite{art20} is obtained to be
\begin{equation}\label{59}
dM = T_{eff} dS_{eff}+V_{eff} dP + \Phi_{eff} dQ ,
\end{equation}
with Eq.\eqref{9} and Eq.\eqref{56} in hand the first law Eq.\eqref{59} becomes
\begin{equation}\label{60}
\begin{split}
dM =& -\dfrac{TT_i}{T+T_i} d(S-S_i)+ \dfrac{VT_i + V_i T}{T+T_i}dP
+\dfrac{\Phi T_i + \Phi_{i} T}{T+T_i}dQ  ,
\end{split}
\end{equation}
with the effective thermodynamic variables given by \cite{Chabab:2020xwr}
\begin{equation}\label{61}
\begin{split}
&T_{eff} = \dfrac{T T_i}{T +T_i}, \hspace{5mm} V_{eff} = \dfrac{V T_i + V_i T}{T +T_i},\\
&\Phi_{eff} = \dfrac{\Phi T_i + \Phi_{i} T}{T +T_i},
\end{split}
\end{equation}
with the effective entropy defined as in \cite{art19,art20},
\begin{equation}\label{62}
S_{eff} = S - S_i
\end{equation}
leading to a vanishing effective entropy for an extremal black hole. 

\begin{figure}[!ht]
	\centering 
	\begin{subfigure}[h]{0.45\textwidth}
		\centering \includegraphics[scale=0.45]{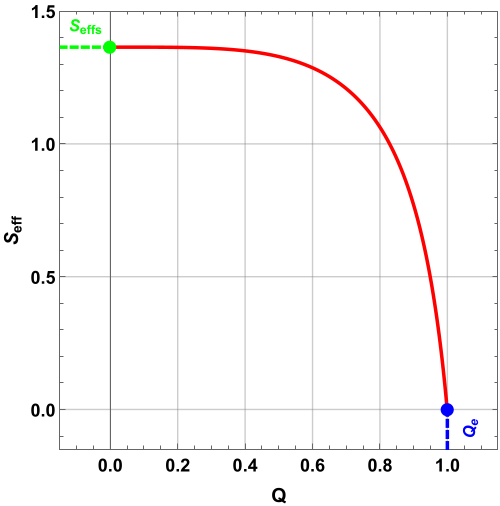}
		\caption{Effective entropy as a function of black hole electric charge.}
		\label{f20_1}
	\end{subfigure}
	\hspace{1pt}	
	\begin{subfigure}[h]{0.45\textwidth}
		\centering \includegraphics[scale=0.45]{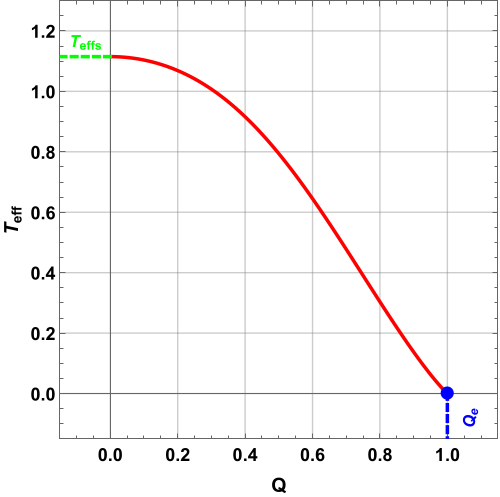}
		\caption{Effective Temperature as a function of black hole electric charge.}
		\label{f20_2}	
	\end{subfigure}
	\hspace{1pt}	
	\begin{subfigure}[h]{0.45\textwidth}
		\centering \includegraphics[scale=0.45]{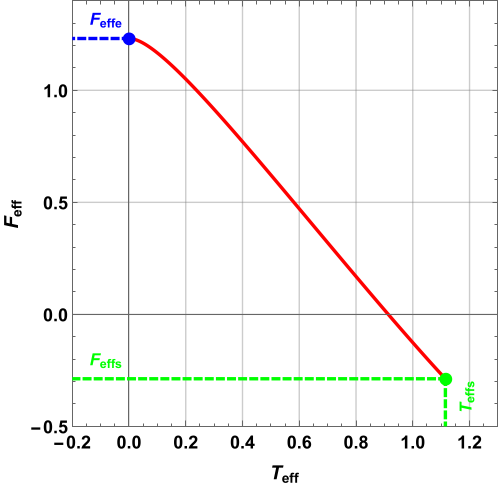}
		\caption{Effective free energy as a function of black hole electric charge.}
		\label{f20_3}	
	\end{subfigure}
	\hspace{1pt}	
	\begin{subfigure}[h]{0.45\textwidth}
		\centering \includegraphics[scale=0.45]{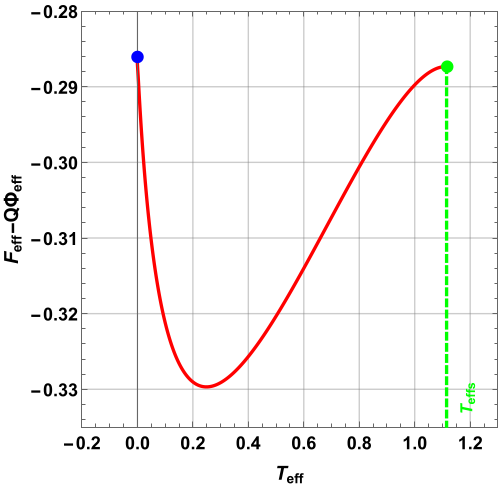}
		\caption{Effective grand potential as a function of black hole electric charge.}
		\label{f20_4}	
	\end{subfigure}
	
	\caption{\footnotesize\it Evolution of different effective thermodynamical variables during an isentropic absorption of $10^3$ particles with a fixed electric charge $dQ = -0.001$. The initial conditions are $Q=1$, $l=1$, $M = 1.23$, $r_h=r_i= 0.65$ and $D=4$. }
	\label{f20}
\end{figure}

Let's now check the validity of the second law within effective entropy, particularly during an isentropic neutralization. Indeed, we have shown that the entropy can be kept constant during an isentropic neutralization which contradicts the fact that the process of absorbing a particle is an irreversible transformation. Then, we illustrate in Fig.~\ref{f20} the behavior of different effective thermodynamic variables during an isentropic neutralization of an extremal black hole in four dimensions.
 From Fig.\ref{f20}, we observe that the effective entropy increases and reaches its maximum at the Schwarzchild-like state even though the entropy associated with the event horizon is constant, the entropy of the inner horizon decreases. This means that this is not a reversible process as the effective entropy is ever-increasing. However, the effective temperature also increases during such a process and has qualitatively the same behavior as the event horizon temperature. The effective free energy shows the same behavior as that of the free energy associated with the event horizon and it increases quasi-linearly in terms of effective temperature to reach its minimum at the Schwarzchild-like state. Whereas the effective grand canonical potential has a different behavior, it grows quickly in terms of effective temperature when the extremal black breaks into a non-extremal one, but once it reaches its minimum it changes its monotony and then increases to have the same value at Schwarzchild-like state as it does in the case of the extremal state. This behavior is not necessarily a sign of phase transition but is due to the change in the electric interaction strength.

\begin{figure}[!ht]
	\centering \includegraphics[scale=0.5]{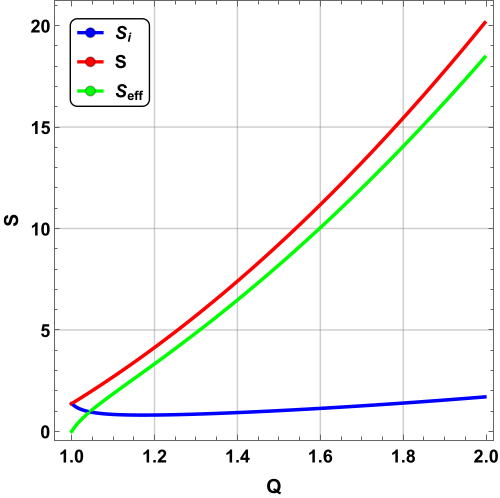}
	\caption{\footnotesize\it Black hole entropy as a function of electric charge when an extremal black hole gets charged absorbing $10^3$ particle with a fixed charge $dQ = 0.001$. The initial conditions are $Q=1$, $l=1$, $M = 1.23$, $r_h=r_i= 0.65$ and $D=4$. }
	\label{f21}
\end{figure}

Similar to the previous analysis, before we finish this subsection we would like to prob the validity of the second law is respected in the effective entropy context when the black bole gets charged. We start with an extremal black hole possessing a charge $+Q$ and then let it get charged until it is doubled in terms of charge. The entropy of such a process is  depicted in Fig.\ref{f21} for a four dimensions case

It can be remarked that both inner horizon entropy and outer entropy (entropy associated with event horizon) increase when the black hole gets charged but the outer entropy increases quicker than the inner horizon entropy and thereby resulting in increasing the effective entropy of the combined inner and outer horizons.
 Therefore, the thermodynamic second law is not only respected by the entropy associated with the event horizon but it is verified and respected also by the effective entropy.

\subsection{Isobar neutralization}

In this last subsection, we will be interested in an isobar neutralization which is to say a neutralization with a fixed pressure. Such a process is very interesting in the context of the thermodynamical description of charged AdS black holes, particularly the first and second-order phase transitions. In addition, such a process also draws our attention to conclude that there exists an equivalence between the charged AdS black hole and Van der Waals-type fluids where the description is generally done at a fixed pressure.

\begin{figure}[h!]
	\centering 
	\begin{subfigure}[h]{0.45\textwidth}
		\centering \includegraphics[scale=0.5]{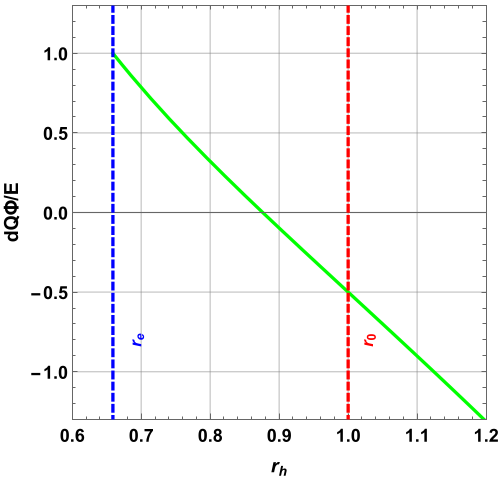}
		\caption{}
		\label{f22_1}
	\end{subfigure}
	\hspace{1pt}	
	\begin{subfigure}[h]{0.45\textwidth}
		\centering \includegraphics[scale=0.5]{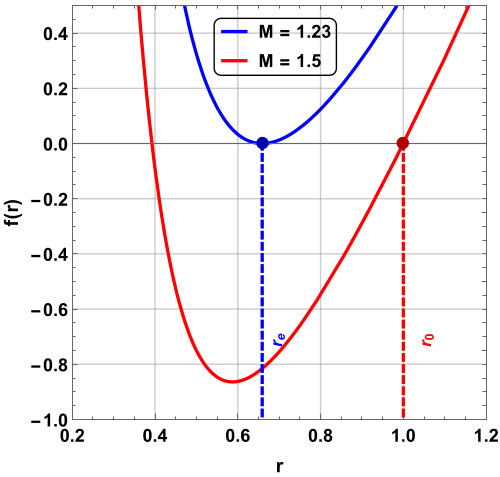}
		\caption{}
		\label{f22_2}	
	\end{subfigure}
	
	\caption{\footnotesize\it (a) Ratio $\dfrac{dQ \Phi}{E}$ and (b) $f(r)$ function for different masses as a function of event horizon radius with $Q=1$, $l=1$ and $D=4$. }
	\label{f22}
\end{figure}

The isobaric absorption can be ensured using Eq.\eqref{30} and then the following constraints 
\begin{equation}\label{63}
\left| p^r_h\right| = - \dfrac{\Phi \left[ \left( D-1\right) r_h^2+\left( D-3\right) l^2\left( 1-\Phi^2\right) \right] }{\left( D-1\right) r_h^2-\left( D-3\right) l^2\left( 1+\Phi^2\right) } dQ,
\end{equation}
where the sign of $dQ$ should be chosen to make $\left| p^r_h\right|$ positive. In addition, we should respect the second condition
\begin{equation}\label{64}
E = \left| p^r_h\right| + dQ\Phi > 0,
\end{equation}
which means that the black hole does not lose energy (nothing can escape from a black hole). Using Eq.\eqref{63} and Eq.\eqref{64}, the condition that the particle should be absorbed with a fixed pressure respecting the condition
\begin{equation}\label{65}
\dfrac{dQ \Phi}{E} < 0,
\end{equation}
which means that an isobar process may occur only when the black gets neutralized. This condition is not always verified particularly for an extremal black hole. Indeed, we see in Fig.\ref{f22} that this condition is not verified for the extremal, thereby we shall study the neutralization of a non-extremal black hole.

\begin{figure}[H]
	\centering 
	\begin{subfigure}[h]{0.45\textwidth}
		\centering \includegraphics[scale=0.5]{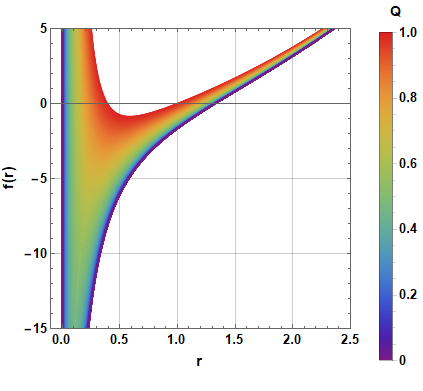}
		\caption{}
		\label{f23_1}
	\end{subfigure}
	\hspace{1pt}	
	\begin{subfigure}[h]{0.45\textwidth}
		\centering \includegraphics[scale=0.5]{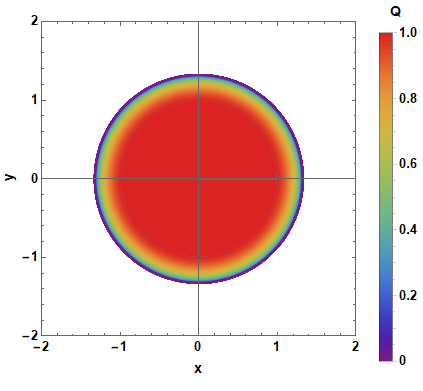}
		\caption{}
		\label{f23_2}	
	\end{subfigure}
	\begin{subfigure}[h]{0.45\textwidth}
		\centering \includegraphics[scale=0.5]{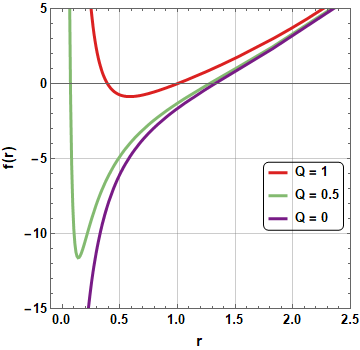}
		\caption{}
		\label{f23_3}
	\end{subfigure}
	\hspace{1pt}	
	\begin{subfigure}[h]{0.45\textwidth}
		\centering \includegraphics[scale=0.5]{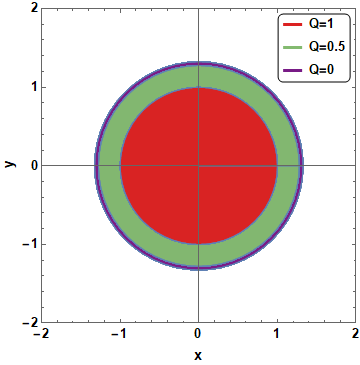}
		\caption{}
		\label{f23_4}	
	\end{subfigure}
	
	\caption{\footnotesize\it (a) and (c): Evolution of $f(r)$ function during absorption of $10^3$ particles with a fixed electric charge $dQ = -0.001$. (b) and (d): Evolution of black hole projection on $xy$ plan during absorption of $10^3$ particles with a fixed electric charge $dQ = -0.001$. The initial conditions are $Q_0=1$, $l=1$, $M_0 = 1.5$, $r_h= 1$ and $D=4$. }
	\label{f23}
\end{figure}

Let's begin the simulation of the absorption process with a non-extremal black hole with $M_0 = 1.5$ and $r_h = 1$ which verifies the condition, Eq.\eqref{65}, as it is shown in Fig.~\ref{f22} and will absorb $10^3$ negatively charged particles with a fixed charge $dQ=-0.001$ in an isobaric process. We show in Fig.~\ref{f23} the evolution of the blackening function $f(r)$ and the black hole projection on $x-y$ plane during the isobar absorption of $10^3$ negatively charged particles.
 We observe that the black hole gets larger quickly when it absorbs the first $500$ particles and then converges slowly to the Scwarzchild-like state.

\begin{figure}[!ht]
	\centering 
	\begin{subfigure}[h]{0.45\textwidth}
		\centering \includegraphics[scale=0.5]{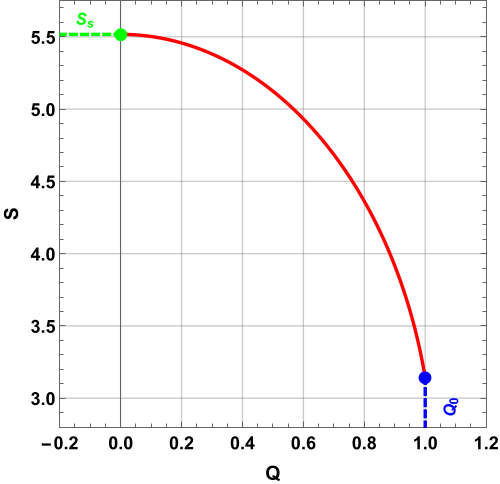}
		\caption{}
		\label{f24_1}
	\end{subfigure}
	\hspace{1pt}	
	\begin{subfigure}[h]{0.45\textwidth}
		\centering \includegraphics[scale=0.5]{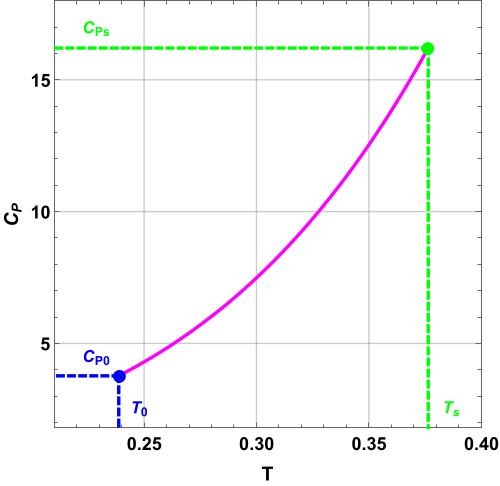}
		\caption{}
		\label{f24_2}	
	\end{subfigure}
	
	\caption{\footnotesize\it (a) Entropy as a function of black hole electric charge during an isobar absorption of $10^3$ particles with a fixed electric charge $dQ = -0.001$. (b) Heat capacity at constant pressure in terms of black hole temperature during an isobar absorption of $10^3$ particles with a fixed electric charge $dQ = -0.001$. The initial conditions are $Q_0=1$, $l=1$, $M_0 = 1.5$, $r_h= 1$ and $D=4$. }
	\label{f24}
\end{figure}

Since the neutralization is isobaric in nature, we can use the heat capacity at a fixed pressure to analyze the stability of the black hole in such a process. We recall that the heat capacity at fixed pressure is given by 
\begin{equation}\label{66}
C_P = \left( \dfrac{\partial M}{\partial T} \right)_P,
\end{equation}
and the system is stable when $C_P > 0$ and unstable  when $C_P < 0$. We graphically reveal in Fig.~\ref{f24} the evolution of entropy and heat capacity at a fixed pressure during the isobar neutralization. 
We observe that the entropy increases when the black hole gets neutralized respecting the second law. The heat capacity curve at fixed pressure shows that the black hole is stable during this process. It is worth noticing that a black could not get charged with an isobar process because such transformation violates the condition Eq.\eqref{65}.

\section{Constant electric potential absorption: Grand canonical description}

To place the last brick of our study,  we will be interested in constant electric potential absorption. The main interest in such a process comes from the context of the thermodynamical description of charged AdS black holes within the grand canonical description.
Unfortunately, we cannot have a neutralization with constant potential because the final state is neutral, hence the potential vanishes. Moreover, we will see below that even partial neutralization is not possible.

To keep the electric potential constant during an absorption we use Eq.\eqref{30} and then the condition reads
\begin{equation}\label{67}
\left| p^r_h\right| =  dQ \dfrac{1-\Phi^2}{\Phi},
\end{equation}
which should be positive and we should respect the second condition given in Eq.\eqref{65}. Combining the two-equation, the conditions to ensure constant electric absorptions are
\begin{equation}\label{68}
\begin{split}
\dfrac{dQ}{\Phi}& > 0, \\
\Phi^2 & < 1,
\end{split}
\end{equation}
hence the constant electric absorption makes the black hole more charged and it is impossible to get neutralized in such a process. 

\begin{figure}[!ht]
	\centering \includegraphics[scale=0.7]{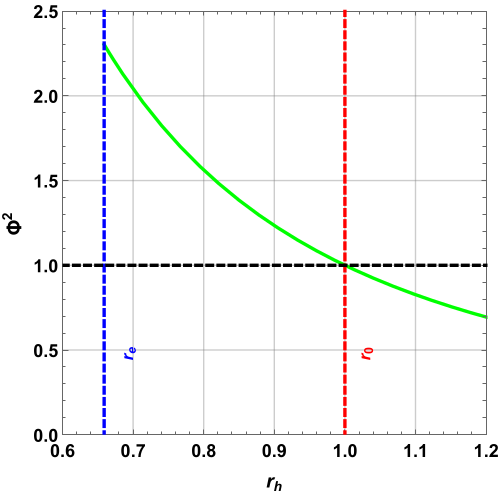}
	\caption{\footnotesize\it Electric potential square as a function of event horizon radius with  $Q=1$, $l=1$, and $D=4$. }
	\label{f25}
\end{figure}

For instance, in four dimensions with $Q=1$, we see in Fig.\ref{f25} 
\clearpage
that a constant electric potential absorption is impossible in the case of extremal and near the extremal situations, but such a process is possible only when $r_h>r_0$. When $r_h = r_0$, the black hole absorbs only the electrostatic energy  ($\left| p^r_h\right|=0$).

\begin{figure}[!ht]
	\centering 
	\begin{subfigure}[h]{0.45\textwidth}
		\centering \includegraphics[scale=0.5]{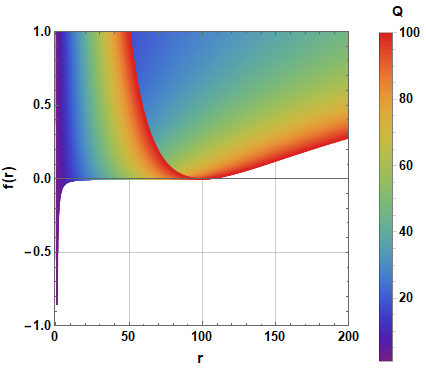}
		\caption{}
		\label{f26_1}
	\end{subfigure}
	\hspace{1pt}	
	\begin{subfigure}[h]{0.45\textwidth}
		\centering \includegraphics[scale=0.5]{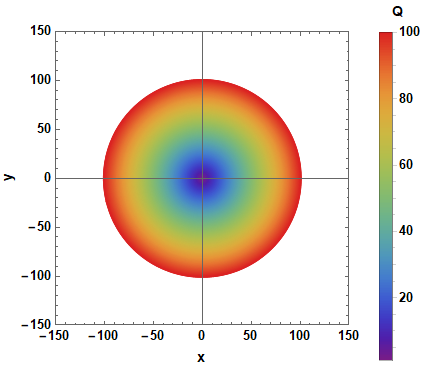}
		\caption{}
		\label{f26_2}	
	\end{subfigure}
	\begin{subfigure}[h]{0.45\textwidth}
		\centering \includegraphics[scale=0.5]{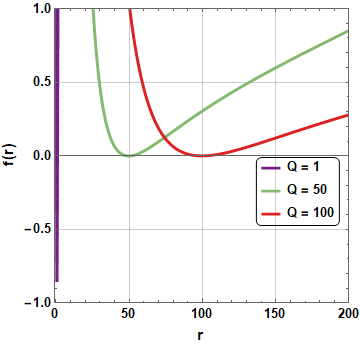}
		\caption{}
		\label{f26_3}
	\end{subfigure}
	\hspace{1pt}	
	\begin{subfigure}[h]{0.45\textwidth}
		\centering \includegraphics[scale=0.5]{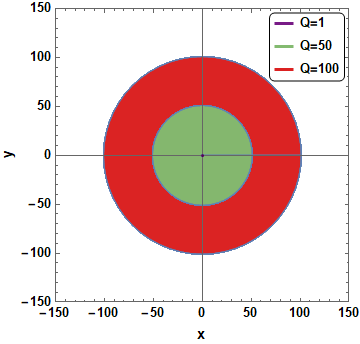}
		\caption{}
		\label{f26_4}	
	\end{subfigure}
	
	\caption{\footnotesize\it (a) and (c): Evolution of $f(r)$ function during absorption of $10^5$ particles with a fixed electric charge $dQ = 0.001$. (b) and (d): Evolution of black hole projection on $xy$ plan during absorption of $10^5$ particles with a fixed electric charge $dQ = 0.001$. The initial conditions are $Q_0=1$, $l=1$, $M_0 = 1.5$, $r_h= 1$ and $D=4$. }
	\label{f26}
\end{figure}

\begin{figure}[!ht]
	\centering 
	\begin{subfigure}[h]{0.45\textwidth}
		\centering \includegraphics[scale=0.5]{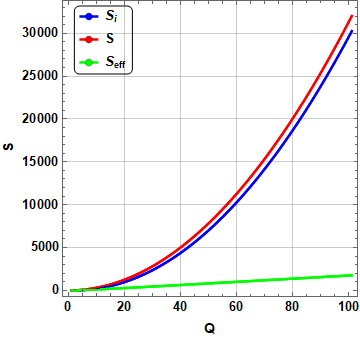}
		\caption{}
		\label{f27_1}
	\end{subfigure}
	\hspace{1pt}	
	\begin{subfigure}[h]{0.45\textwidth}
		\centering \includegraphics[scale=0.5]{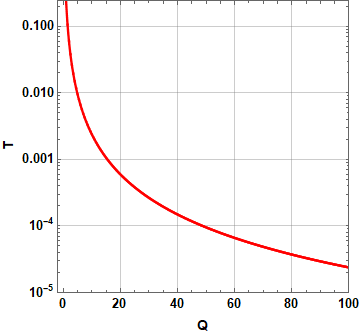}
		\caption{}
		\label{f27_2}	
	\end{subfigure}
	
	\caption{\footnotesize\it (a) Entropy as a function of black hole charge during absorption of $10^5$ particles with a fixed electric charge $dQ = 0.001$. (b) $f_{min} = f(r_{min})$ in terms of black hole electric charge during absorption of $10^5$ particles with a fixed electric charge $dQ = 0.001$. The initial conditions are $Q_0=1$, $l=1$, $M_0 = 1.5$, $r_h= 1$ and $D=4$. }
	\label{f27}
\end{figure}

We begin with a non-extremal black hole with $r_h = r_0$, $Q=1$, and $l=1$ that will absorb $10^5$ particles with $\left| p^r_h\right|= 0$ such that the electric potential is kept constant. We plot again in Fig.\ref{f26} the evolution of  $f(r)$ function and the black hole projection on $xy$ plan during the absorption of $10^5$ positive particles.
 We observe that when the black hole absorbs particles at the constant electric potential it gets larger but it tends to form an extremal black hole. Therefore, one verifies the second law as we can observe in Fig.~\ref{f27} that the entropy of the black (associated with the event horizon) increases during the absorption.
 Moreover, the inner entropy increases also such that the effective entropy given in Eq.\eqref{62} increases too. Hence, the evolution of effective entropy verifies the second law as well. We notice that at the extremal state, the effective entropy vanishes, but as it increases during the absorption, leading to conclude that we cannot reach an extremal state even though $f_{min} = f(r_{min})$ tends to zero (but it is always different from zero !) as it is shown in Fig.\ref{f27}.
 This result verifies and respects the third law of thermodynamics which stipulates that it is impossible for any process, no matter how idealized, to reduce the entropy of a system to its absolute-zero value in a finite number of operations.

\begin{figure}[!ht]
	\centering 
	\begin{subfigure}[h]{0.45\textwidth}
		\centering \includegraphics[scale=0.5]{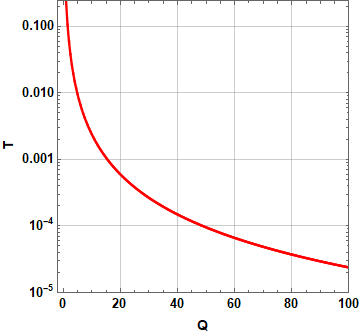}
		\caption{}
		\label{f28_1}
	\end{subfigure}
	\hspace{1pt}	
	\begin{subfigure}[h]{0.45\textwidth}
		\centering \includegraphics[scale=0.5]{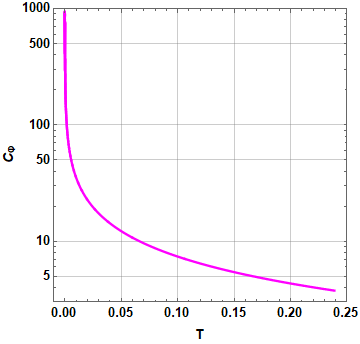}
		\caption{}
		\label{f28_2}	
	\end{subfigure}
	
	\caption{\footnotesize\it (a) Temperature as a function of black hole charge during absorption of $10^5$ particles with a fixed electric charge $dQ = 0.001$. (b) Heat capacity at the constant electric potential in terms of black hole temperature during absorption of $10^5$ particles with a fixed electric charge $dQ = 0.001$. The initial conditions are $Q_0=1$, $l=1$, $M_0 = 1.5$, $r_h= 1$ and $D=4$. }
	\label{f28}
\end{figure}

We illustrate in Fig.\ref{f28} the evolution of the black hole temperature as a function of the electric charge 
and we  observe that it decreases quickly and tends to zero (but it is always different from zero !), thereby the system tends to form an extremal black hole but it is possible at a finite charge, it can be formed only with an infinite charge confirming to the third law of thermodynamics. As the absorption occurs at fixed electric potential, we define the heat capacity at the fixed electric potential as
\begin{equation}\label{69}
C_{\Phi} = \left(T \dfrac{\partial S}{\partial T} \right)_{\Phi},
\end{equation}
and the system criteria of stability is controlled by the sign of the heat capacity, indeed the system is stable when $C_{\Phi} > 0$ and unstable  when $C_{\Phi} < 0$. 
 The heat capacity at fixed electric potential in terms of black hole temperature is also depicted in the right panel of Fig.\ref{f28},
and from which, we notice that $C_{\Phi}$ is always positive,  meaning that the black hole is stable during such absorption.

\begin{figure}[H]
 	\centering 
 	\begin{subfigure}[h]{0.45\textwidth}
 		\centering \includegraphics[scale=0.5]{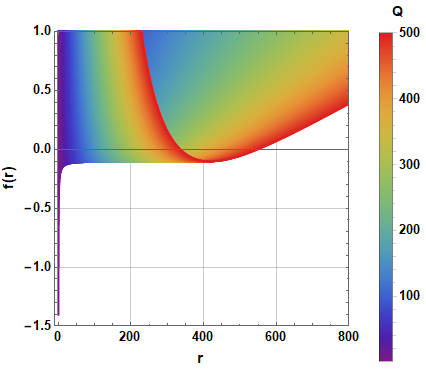}
 		\caption{}
 		\label{f29_1}
 	\end{subfigure}
 	\hspace{1pt}	
 	\begin{subfigure}[h]{0.45\textwidth}
 		\centering \includegraphics[scale=0.5]{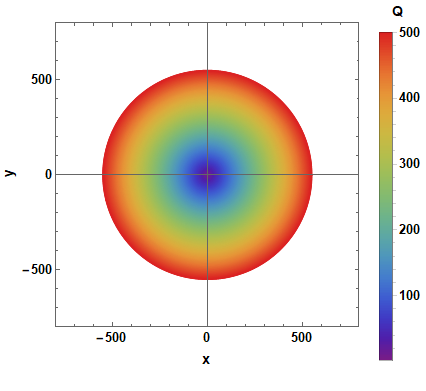}
 		\caption{}
 		\label{f29_2}	
 	\end{subfigure}
 	\begin{subfigure}[h]{0.45\textwidth}
 		\centering \includegraphics[scale=0.5]{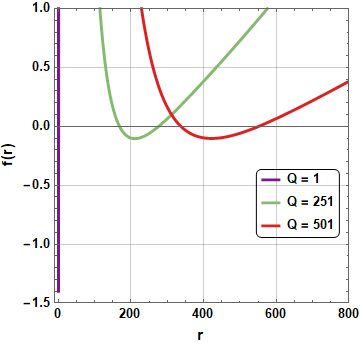}
 		\caption{}
 		\label{f29_3}
 	\end{subfigure}
 	\hspace{1pt}	
 	\begin{subfigure}[h]{0.45\textwidth}
 		\centering \includegraphics[scale=0.5]{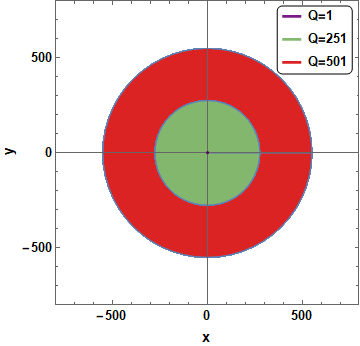}
 		\caption{}
 		\label{f29_4}	
 	\end{subfigure}
 	
 	\caption{\footnotesize\it (a) and (c): Evolution of $f(r)$ function during absorption of $10^5$ particles with a fixed electric charge $dQ = 0.005$. (b) and (d): Evolution of black hole projection on $xy$ plan during absorption of $10^5$ particles with a fixed electric charge $dQ = 0.005$. The initial conditions are $Q_0=1$, $l=1$, $M_0 = 1.67$, $r_h= 1.1$ and $D=4$. }
 	\label{f29}
 \end{figure}

\begin{figure}[!ht]
	\centering 
	\begin{subfigure}[h]{0.45\textwidth}
		\centering \includegraphics[scale=0.5]{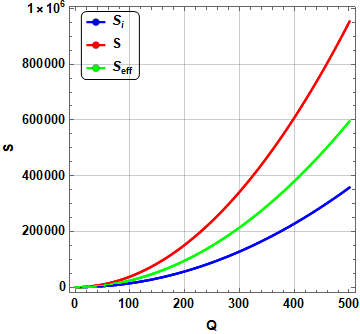}
		\caption{}
		\label{f30_1}
	\end{subfigure}
	\hspace{1pt}	
	\begin{subfigure}[h]{0.45\textwidth}
		\centering \includegraphics[scale=0.5]{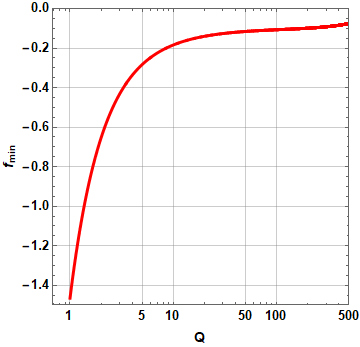}
		\caption{}
		\label{f30_2}	
	\end{subfigure}
	
	\caption{\footnotesize\it (a) Entropy as a function of black hole charge during absorption of $10^5$ particles with a fixed electric charge $dQ = 0.005$. (b) $f_{min} = f(r_{min})$ in terms of black hole electric charge during absorption of $10^5$ particles with a fixed electric charge $dQ = 0.005$. The initial conditions are $Q_0=1$, $l=1$, $M_0 = 1.67$, $r_h= 1.1$ and $D=4$. }
	\label{f30}
\end{figure}

In order to verify that this behavior is the same for all constant potential processes we take another scenario where  $\left| p^r_h\right|\neq 0$. From Fig.\ref{f25} we choose $r_h = 1.1$ as an initial condition and we make the black hole absorb $10^5$ particles according to Eq.\eqref{67}, then, 
 we plot in Fig.\ref{f29} the evolution of  $f(r)$ function and the black hole projection on $xy$ plan during the absorption of $10^5$ positive particles.
One can see that the black hole has qualitatively the same behavior as in Fig.\ref{f26} but it needs to absorb a great charge to tend to an extremal state. Indeed, when it absorbs $\Delta Q = 500$ it is still far from the extremal state. 
Hence, one checks the second law as we can observe in Fig.\ref{f30} that the entropy of the black (associated with the event horizon) increases during the absorption. 
Moreover, the inner horizon entropy increases also such that the effective entropy given in Eq.\eqref{62} increases too. Hence, the evolution of effective entropy verifies the second law as well. We notice also that at the extremal state, the effective entropy vanishes, but as it increases during the absorption we conclude that we cannot have an extremal state even though $f_{min} = f(r_{min})$ tends to zero (but it is always different from zero !) as it is shown in Fig.\ref{f30}. This result is in good agreement with the third law of thermodynamics.

\begin{figure}[H]
	\centering 
	\begin{subfigure}[h]{0.45\textwidth}
		\centering \includegraphics[scale=0.5]{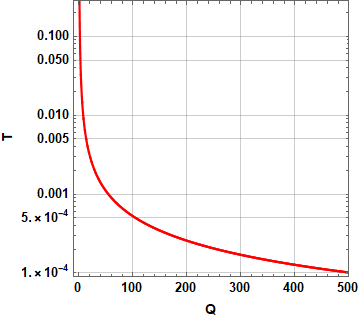}
		\caption{}
		\label{f31_1}
	\end{subfigure}
	\hspace{1pt}	
	\begin{subfigure}[h]{0.45\textwidth}
		\centering \includegraphics[scale=0.5]{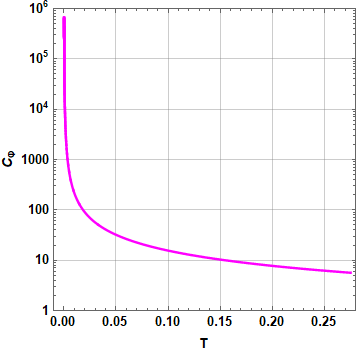}
		\caption{}
		\label{f31_2}	
	\end{subfigure}
	
	\caption{\footnotesize\it (a) Temperature as a function of black hole charge during absorption of $10^5$ particles with a fixed electric charge $dQ = 0.005$. (b) Heat capacity at the constant electric potential in terms of black hole temperature during absorption of $10^5$ particles with a fixed electric charge $dQ = 0.005$. The initial conditions are $Q_0=1$, $l=1$, $M_0 = 1.67$, $r_h= 1.1$ and $D=4$. }
	\label{f31}
\end{figure}

We plot in Fig.\ref{f31} the evolution of the black hole temperature as a function of the electric and the heat capacity at fixed electric potential in terms of black hole temperature. We  observe that it decreases quickly and tends to zero (but it is always different from zero !), thereby the system tends to form an extremal black hole but it is possible at a finite charge, it can be formed only with an infinite charge confirming to the third law of thermodynamics. Moreover, $C_{\Phi} > 0$ which means that the black hole is stable during such absorption.

Further,  in Fig.\ref{f31} we give the variation of the black hole temperature as a function of the electric charge (left panel) and the heat capacity at fixed electric potential in terms of black hole temperature (right panel).
 We  observe that it decreases quickly and tends to zero (but it is always different from zero !), thereby the system tends to form an extremal black hole but it is possible at a finite charge, it can be formed only with an infinite charge confirming to the third law of thermodynamics. Moreover, $C_{\Phi} > 0$ means that the black hole is stable during such absorption.

\begin{figure}[H]
	\centering 
	\begin{subfigure}[!ht]{0.45\textwidth}
		\centering \includegraphics[scale=0.5]{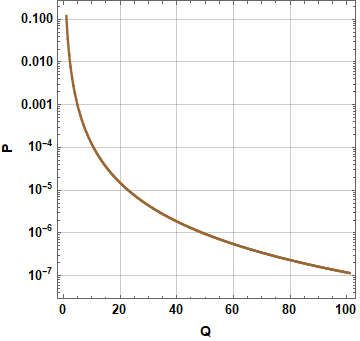}
		\caption{$M_0 = 1.5$ and $r_h= 1$.}
		\label{f32_1}
	\end{subfigure}
	\hspace{1pt}	
	\begin{subfigure}[!ht]{0.45\textwidth}
		\centering \includegraphics[scale=0.5]{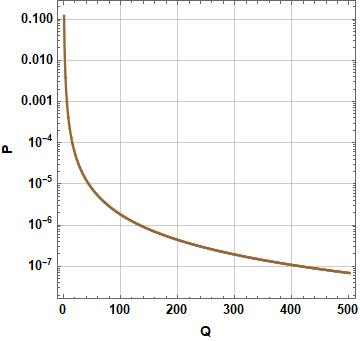}
		\caption{$M_0 = 1.67$ and $r_h= 1.1$.}
		\label{f32_2}	
	\end{subfigure}
	
	\caption{\footnotesize\it Black hole pressure as a function of electric charge during a constant electric potential absorption. The initial conditions are $Q_0=1$, $l=1$, and $D=4$. }
	\label{f32}
\end{figure}

Finally, we depict in Fig.\ref{f32} the evolution of black hole pressure as a function of the electric charge for different initial conditions.
 We observe that the pressure decreases during the absorption, that is to say, the cosmological constant (which is negative) increases, thus the AdS spacetime tends to be a flat spacetime.

\section{Conclusion}
The black hole thermodynamics in AdS spacetime, where the mass $M$ is considered to be enthalpy, has been brought to the forefront of modern theoretical high-energy physics, because of its relevance to the AdS/CFT correspondence. Promoting the status of the negative cosmological constant ($\Lambda<0$), as a thermodynamic pressure, $P=-{\Lambda}/{8\pi}$, and defining the thermodynamic volume, $V=\frac{\partial M}{\partial P}|_{Q,r_h}$, have enriched the thermodynamics analysis of the phase structure of the AdS black holes. In this work, we revisit the second law and the WCCC for the higher dimensional charged AdS black holes, where the absorption of the charged scalar particles took place by the black hole itself. The absorption of the charged particles alters the initial configurations of the black holes, and therefore the final state might transit to another black hole configuration. Assuming the final configuration as the black hole, we verify that the first law still holds even if it absorbs charged particles. We find that the minimum of the blackening function $f(r)$ got shifted to the left or the right depending on the charged particles' absorption by the black hole. We consider both the extremal and near extremal configurations of the black holes in order to verify the possible existence of the black holes' event horizon and consequently the WCCC to be held. The justification was made by evaluating the blackening factor $f(r)$. These arguments validated the WCCC through the investigation of the function $f(r)$ around the minimum point. We started with the extremal and near extremal conditions at the initial stage, and end with the different extremal and near extremal black hole configurations at the final state with the proviso that the condition $dr_{\text{min}}<dr_h$ is held. But for $dr_{\text{min}}=dr_h$ the initial and final extremal configurations are the same. As a generic condition, if a charged scalar field was getting swallowed up by the black hole it would change the location of the minimum and the variation of the minimum point has been plotted as a function of the spacetime dimension $D$. Though the minimum point is not fixed for the initial and final non-extremal situations, still it preserves the existence of the event horizon, thereby respecting the WCCC throughout the process.\\

To this end, perhaps the most important analysis is the black hole neutralization and the validation of the second law. We investigated the neutralization process at constant black hole's event horizon entropy (isentropic process), and also through the effective entropy formalism. We observed that the second law always holds to be good as the entropy always increases through an irreversible process. It has also a relation connecting the electric potential, the electric charge, and the AdS radius as well as the variations of electric charge and the AdS radius. When we keep the value of the AdS radius fixed and vary the electric charge we found for different choices of the electric charges gave different results. Similarly, for a fixed electric charge, if we vary the AdS radius then also we saw that it would give rise to different results. We also envisaged the isobar neutralization to verify the first and second-order phase transitions of the charged AdS black holes in higher dimensions. In this process also, the entropy increases when the black hole got neutralized and thereby respecting the universality of the second law. We can speculate from these analyses that for the (non-)extremality conditions the overcharging of the black holes is not possible as it would lead to the violation of the universality of the second law.\\

An important related analysis is that of constant electric potential in the context of extended-phase space thermodynamics. Such a grand canonical description was made to give a final touch to our study toward verifying the second law of black hole mechanics. Another interesting result that we could infer from such a study is the verification of the third law of the black hole mechanics, i.e., to say during the absorption of the constant electric potential the effective entropy increases and we could never reach the extremal state even though $f(r)$ tends to zero at $r=r_{\text{min}}$. We also observed that the event horizon temperature as a function of charge and the heat capacity as a function of the temperature approached the near extremal configurations. But they couldn't approach an extremal state for a finite charge value of the positively charged particles. It could be attained at the charge value $Q\to\infty$ and thereby confirm the third law of black hole mechanics. We also inferred from the heat capacity $C_\Phi$ that it is always positive, and hence the black hole is always thermodynamically stable.\\

In the context of the AdS/CFT correspondence, the contribution of a negative cosmological constant in the first law of the black holes' AdS bulk is equivalent to the variation of the number of degrees of freedom, $N$, of the boundary field theory CFT. We wish to study our analysis in terms of the CFT dictionary, where the first law should include the chemical potential term. We shall return to this issue in the near future.

\section*{Acknowledgment}
The research of M. S. A. is supported by the National Postdoctoral Fellowship of the Science and Engineering Research Board (SERB), Department of Science and Technology (DST), Government of India, File No., PDF/2021/003491.

\bibliographystyle{unsrt}
\bibliography{slwccc.bib}
\end{document}